%% file: main-paper.tex
\documentclass[review]{elsarticle}

\usepackage{lineno,hyperref}
\modulolinenumbers[5]

\journal{arXiv}

\usepackage{url}

\PassOptionsToPackage{table,xcdraw}{xcolor}

\input{addons/pgload}

\input{addons/glossary}

\begin{document}

\begin{frontmatter}

\title{A Survey of Approaches for Event Sequence Analysis and Visualization using the ESeVis Framework} 

\author[1]{Anton Yeshchenko}

\address[1]{Vienna University of Economics and Business, Vienna, Austria}

\author[2]{Jan Mendling}

\address[2]{Humboldt-Universit{\"a}t zu Berlin, Berlin, Germany}

%
%
%

\begin{abstract}
\input{sections/abstract}
\end{abstract}


\end{frontmatter}


\newpage


%
%
\section{Introduction}
\label{sec:intro}
\input{sections/intro}

%
%
%
%
%
\section{Background}
\label{sec:background}
\input{sections/background}
\section{Research design}
\label{sec:research_design}
\input{sections/research-design}

%
%
%

%
\section{Findings}
\label{sec:results}
\input{sections/results}
\section{Conclusions}
\label{sec:conclusions}
\input{sections/conclusions}
\section*{Acknowledgement}
\noindent The research by Jan Mendling was supported by the Einstein Foundation Berlin.

\bibliography{library}

\end{document}

%% file: addons/pgload.tex
\usepackage[textsize=scriptsize,backgroundcolor=yellow!40]{todonotes}
\usepackage{graphicx,wrapfig,lipsum}

\usepackage[nonumberlist,acronym,sanitize=none]{glossaries}

\usepackage{booktabs}

\usepackage{booktabs}
\usepackage{multirow}
\usepackage{rotating}
\usepackage{siunitx}
\usepackage{adjustbox}
\usepackage{wrapfig}

\usepackage{subfig}

\usepackage{listings}

\usepackage{cleveref}

\usepackage{graphbox}

\usepackage{makecell}

\usepackage{tabularx}

%% file: addons/glossary.tex
\newacronym{ex-data}{Agriculture Fund}{European Agricultural Guarantee Fund}

\newacronym{is}{IS}{Information Systems}
\newacronym{tkde}{TKDE}{IEEE Transactions on Knowledge and Data Engineering}
\newacronym{dss}{DSS}{Decision Support Systems}
\newacronym{tvcg}{TVCG}{IEEE Transactions on Visualization and Computer Graphics}
\newacronym{iv}{IV}{Information Visualization}
\newacronym{cgf}{CGF}{Computer Graphics Forum}

\def\timelineA {Fixed timeline-based}
\def\timelineB {Duration timeline-based}
\def\timelineC {Converging-diverging timeline-based}
\def\timelineD {Evolution timeline-based}
\def\timelineE {Combinations}

\def\hierarchyA {Node-link hierarchy-based}
\def\hierarchyB {Flow hierarchy-based}
\def\hierarchyC {Treemap hierarchy-based}

\def\matrixA {Matrix-based}
\def\matrixB {Extended Matrix-based}

\def\sankeyA {Node-link sankey-based}
\def\sankeyB {Extended node-link sankey-based}
\def\sankeyC {Flow sankey-based}
\def\sankeyD {Converging-diverging  sankey-based}
\def\sankeyE {Chord  sankey-based}

\def\chartA {Sequence chart-based}
\def\chartB {Sequence-duration chart-based}
\def\chartC {Bar-chart-based}

%% file: sections/abstract.tex
Event sequence data is increasingly available. Many business operations are supported by information systems that record transactions, events, state changes, message exchanges, and so forth. This observation is equally valid for various industries, including production, logistics, healthcare, financial services, education, to name but a few. The variety of application areas explains that techniques for event sequence data analysis have been developed rather independently in different fields of computer science. Most prominent are contributions from information visualization and from process mining. So far, the contributions from these two fields have neither been compared nor have they been mapped to an integrated framework. In this paper, we develop the Event Sequence Visualization framework (ESeVis) that gives due credit to the traditions of both fields. Our mapping study provides an integrated perspective on both fields and identifies potential for synergies for future research.

%% file: sections/intro.tex
\noindent
Event sequence analysis is an important field of computer science due to its relevance to a diverse spectrum of application domains such as manufacturing, logistics, healthcare, financial services, education~\citep{dos2019process}, to name but a few.
Despite this broad relevance across these domains, it is striking to observe that techniques for event sequence data analysis have been developed rather independently in different fields of computer science. 

The most prominent research fields investigating the analysis of event sequence data are process mining and information visualization. 
\emph{Process mining} has emerged as a subfield of research into workflow management systems~\citep{DBLP:books/sp/Aalst16}. Its focus is the development of new techniques for automatic process discovery from event sequence data with the ambition to provide a meaningful and understandable summary of the behaviour to the business process analyst. 
\emph{Information visualization} is a field of computer graphics, which originated as a subfield of human-computer interaction~\citep{DBLP:series/hci/AignerMST11}. Its focus is on devising new techniques for visualizing event sequence data in a meaningful way such that analysts can effectively explore them. Typical representations frequently used in this field are timelines that plot conceptually related sequences of events over a time axis.
As similar as the ambitions of these research areas may sound, it is surprising that there is hardly any exchange of ideas. Cross-references are scarce and mutual awareness and understanding is limited.\footnote{For instance, by December 2021 there are less than ten articles in IEEE Transactions on Visualization and Computer Graphics that mention the term ``process mining'' at all.} All this makes research on event sequence analysis a fragmented field with scattered contributions.

So far, the contributions from these two fields have neither been compared nor have they been mapped to an integrated framework. For this reason, it is not clear to which extent both fields have developed complementary concepts and insights.
In this paper, we develop such a framework that we call EseVis and that gives due credit to the traditions of both fields. Our mapping study provides an integrated perspective on both fields and potential synergies for future research. In this way, our work contributes towards overcoming the fragmentation of research on event sequence data analysis.

The paper is structured as follows. Section~\ref{sec:background} discusses the background of our research. We use the example of event sequence data of sepsis treatments in a hospital to illustrate and define categories of event sequence data representations and visualizations. Based on both representations and visualizations, we develop our ESeVis framework. Section~\ref{sec:research_design} describes our method of systematically reviewing the literature on event sequence data analysis in the fields of information visualization and process mining. Section~\ref{sec:results} presents our findings. We provide descriptive results on the distribution of contributions over the different categories of the ESeVis framework. Furthermore, we illustrate key concepts for each category of the framework. Section~\ref{sec:conclusions} concludes the paper and gives an outlook on future research.

%% file: sections/background.tex
\noindent
In this section, we discuss the background of our research. First, we describe the various types, characteristics and formats of event sequence data. Then, we distinguish two distinct conceptual categories of event sequence representation, namely sequence representation and model representation. Moreover, we identify five visualization techniques. Based on conceptual representation and visualization technique, we develop our ESeVis framework for event sequence data analysis. We use the term representation to refer to the formal syntax and semantics of the event sequence data types and visualization to refer to specific visualization techniques, similar to the distinction made by Keim~\cite{keim2002information}.

\subsection{Example of Sepsis Treatment in a Hospital}
\noindent
Event sequence data originates from a variety of real-world processes. For example, such data is recorded by information systems, such as enterprise resource planning systems~\cite{DBLP:books/sp/DumasRMR18} or healthcare systems~\cite{DBLP:journals/tvcg/BernardSKR19}). These systems capture sequences of transactions, web click streams~\cite{DBLP:journals/tvcg/NguyenTAATZ19}, student learning progress~\cite{DBLP:journals/cgf/SungHSCLW17}, or any other traces of human behavior~\cite{DBLP:journals/tvcg/VrotsouJC09,DBLP:journals/tvcg/JangER16}. The common characteristic of these event sequence data is that they encode sequences of events over time. 

We use the example of the event sequence data originating from the transaction system of a hospital for illustrating important characteristics of such data. This example covers the event sequences of handling \emph{sepsis} cases~\cite{DBLP:conf/emisa/MannhardtB17}. The events of these sequences refer to the progression of patients in the hospital including registration, tests, diagnoses, possible transfer to care units, and release. \Cref{table:sepsis-typical-example} summarizes the most frequent steps of this process. These represent the typical treatment sequence of a patient.

\begin{table} 
	\caption{Example of a typical steps in the Sepsis process based on data of~\cite{DBLP:conf/emisa/MannhardtB17}.}
	\label{table:sepsis-typical-example}
	\centering
	\small
		\input{tables/example-sepsis.tex}

\end{table}


\subsection{Representation of Events in Event Sequence Data} 
\label{subsec:event-seq-data}
\noindent
Event sequence data represents event sequences such as documented cases of treating sepsis patients. An event sequence is an ordered list of events~\cite{DBLP:journals/sigkdd/XingPK10}. One event represents a step in a process, such as \emph{ER Sepsis Triage} for the \emph{Sepsis} process. Each sequence has a start event, such as \emph{ER Registration}, and finishes with an end event, such as \emph{Release A}. Events can be represented as a symbolic value, numerical value, or as a complex data type comprising of a set of values of different types. It depends on the real-world phenomenon and the supporting systems which kind of event data can be recorded. 
In particular, we distinguish different categories of event data~\cite{DBLP:journals/sigkdd/XingPK10}. 


\begin{itemize}
	\item{\emph{Simple symbolic sequence:}} If each event in the sequence can be described by one categorical value, then the sequence of such values forms a simple symbolic sequence. Formally, this data is an ordered list of events instantiated from some alphabet \{$e_1,e_2,\dots,e_n$\}. These event sequences can represent web click streams or program execution sequences. The sequence of sepsis-related treatment events for one patient are, for instance, \emph{ER Registration}, \emph{ER Triage}, \emph{LacticAcid}, \emph{Release A}.
	
	\item{\emph{Complex symbolic sequence:}} When events are represented as tuples of categorical values, the event sequences formed by such events is called complex symbolic sequence. This type of data is an ordered list of vectors that capture values drawn from some alphabet \{$\vec{e_1},\vec{e_2},\dots,\vec{e_n}$\}. 
	One example of recording events related to procedures performed on a patient for~\emph{Sepsis} case would be: \\ $\langle(ER Registration, Private Insurance), \\(ER Triage, Dr. Garcia), \\(LacticAcid, Leucocytes, Dr. Williams),\\ ..., (Release A) \rangle$. 
	
	\item{\emph{Simple time series:}} If an event represents a single numerical value combined with the time when this value was recorded, the event sequences originating from these events is called a simple time series. This type of data is a sequence of timestamped real-valued numbers, representing a change of a value over predefined time periods. 
	In the \emph{Sepsis} process, the time series reflecting the current number of treated patients in the clinic is one example 
	and another one the change of health indicators of a patient over time. For instance, $\langle(t_1, 37^{\circ}), (t_2, 39^{\circ}), ..., (t_n, 38^{\circ}) \rangle$ can represent the body temperature of a patient. 
	
	\item{\emph{Multivariate time series:}} In this representation, each event is captured as several numerical values. This type of data is a sequence of timestamped real-valued vectors, describing how these vectors change over time. Each single event is represented by a list of values. In the \emph{Sepsis} example, several health statistics related to a patient can be grouped together into multivariate time series. The change of body temperature and blood sugar can be represented as multivariate time series $\langle(t_1,\langle 37.9^{\circ}, 190\rangle), ..., (t_n,\langle 36.6^{\circ}, 120\rangle) \rangle$
	
	
	\item{\emph{Complex event sequence:}} When real-world phenomena are captured with rich information, recorded events can contain complex content. This type of data is the sequence of values that have complex data types, each event containing a number of values, usually including timestamp and multiple other event characteristics of textual, numerical and categorical types. For each event related to a patient in the \emph{Sepsis} data set, the timestamp, event name, and many other characteristics can be recorded. The example of an event can be 10:05 10.05.2021$\langle Admission IC, 36.6^{\circ}, 120\% \rangle$ that describes the event \emph{Admission IC} that was recorded on 10:05 10.05.2021, and the person in question had a body temperature of 36.6 and a sugar level at 120.

	
\end{itemize}



The choice of \emph{how} events are represented in event sequence data determines which analysis can be done. Simple symbolic sequence and complex symbolic sequences can be used to discover the sequential patterns of events. The simple and multivariate time series can be used to calculate regression and numerical statistics. The complex event sequences can be used for a combination of these analyses.

\subsection{Representation of Event Sequences}
\noindent
No matter how events are stored, we can distinguish two ways how event sequences can be represented. First, we describe different \emph{instance-based representations} of event sequences. These representations have in common that they list event sequences according to different principles. They describe the actual event sequences that have been observed. Second, \emph{model-based} representations do not describe individual sequences, but the rules by which these sequences can be generated. They abstract from actual event sequences. We describe both representations in turn.

\subsubsection{Instance-based Representation of Event Sequences}
\begin{figure}[h]
	\centering
	\resizebox{\columnwidth}{!} {
		\includegraphics[width=\linewidth]{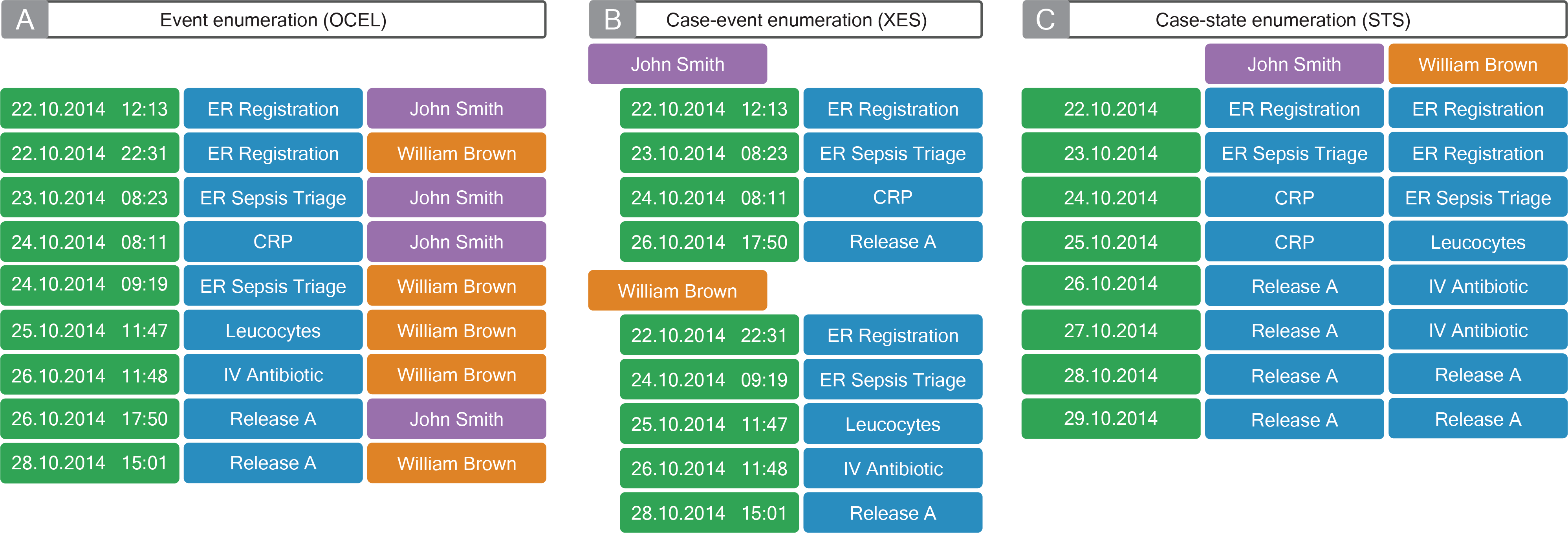}}
	\caption{Examples of different instance-based representations of event sequences}
	\label{fig-def-xes-ocel-sts}
\end{figure}

\noindent
Instance-based representations organize data on events of several event sequences in different ways. We distinguish event enumeration, case-event enumeration, and case-state enumeration (see Figure~\ref{fig-def-xes-ocel-sts}). Different file formats exist for each of these categories.

%

\paragraph{Event enumeration} This event sequence representation stores data as a sequence of events. These events contains the event name, timestamp, and any additional attributes related to the event. Additional attributes contain important information that can be used to order events into separate event sequences (or cases). One example for event enumeration is the \emph{OCEL file format} that is used to store lists of events and support multiple case notions~\cite{ocelstandard}. An example of data stored in this format is shown in~\Cref{fig-def-xes-ocel-sts}.A. The OCEL format was developed to solve the problem of storing events that can be grouped according to different case notions into different event sequences. For instance, events of the sepsis data collection can be grouped into sequences based on the same patient (sequence of the treatment) or based on the same nurse (sequence of a shift).
OCEL was designed to use XML-based syntax to store event sequence data with a possibility of storing also the information about the relation between objects within event sequences. That information allows the data analysis to identify the resulting case notion and resulting event sequences at the stage of analysis. OCEL provides a general process mining standard to interchange object-centric event data with multiple case notions.


\paragraph{Case-event enumeration} This type of storing event sequences requires the definition of an order of events and their reference to one specific \emph{case}. This yields a hierarchical representation, where each event belongs to a case, which usually represents one instance of a process. The \emph{eXtensible Event Stream (XES) format} is an XML-based standard event sequence format for information systems~\cite{xes2016citation}. This format is used by software tools for process mining, such as ProM~\cite{DBLP:conf/apn/DongenMVWA05} or Disco~\cite{disco}.
%
%
An example of storing data in this way is shown in~\Cref{fig-def-xes-ocel-sts}.B. It covers the Sepsis data in the \emph{XES} format. Using case-event enumeration, each event is assigned to a case. Here, each case represents one patient in the clinic, and the events represent the procedures performed.


\paragraph{Case-state enumeration} This representation focuses on the event sequences as changes of states for a particular case. The \emph{STate Sequence (STS)} format considers states as a data unit instead of events~\cite{gabadinho2011analyzing}. This representation still uses the tabular format to store data, but with a different structure. As shown in~\Cref{fig-def-xes-ocel-sts}.C, rows correspond to the sequence of equitemporally spaced timestamps with corresponding states for each case in that time.

\subsubsection{Model-based Representation of Event Sequences}
\noindent
Model-based representations (or for short: models) describe how individual sequences can be generated and which characteristics they have. In this way, they aggregate event sequences into abstract specifications. 
Several categories of the model-based representations of event sequences exist. We describe four examples that are frequently used for process mining, namely directly-follows graphs, Petri nets, Declare constraints, and process trees~\citep{DBLP:books/sp/Aalst16}. It is well known that transformations exist between many of these representations~\citep{de2007use,DBLP:journals/jcss/DeutchM12,DBLP:conf/simpda/PrescherCM14}

\begin{figure}[h]
	\centering
	\resizebox{0.8\columnwidth}{!} {
		\includegraphics[width=\linewidth]{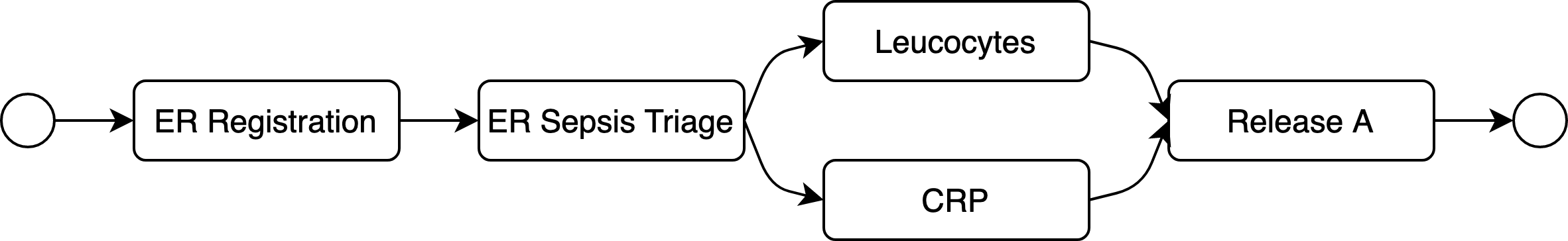}}
	\caption{Directly-follows graph for the 
		\emph{Sepsis} example}
	\label{fig-dfg-sepsis}
\end{figure}

\paragraph{Directly-follows graphs} Event sequences can be modeled as a graph, where each node represents a type of event and arcs describe directly-follows relationships between these event types~\cite{DBLP:books/sp/Aalst16}. An exemplary directly-follows graph using Sepsis is shown in~\Cref{fig-dfg-sepsis}. In essence, it shows for each event types like \emph{ER Sepsis Triage} the event types that can follow next in an event sequence, here \emph{Leucocytes} and \emph{CRP}.

\begin{figure}[h]
	\centering
	\resizebox{0.7\columnwidth}{!} {
		\includegraphics[width=\linewidth]{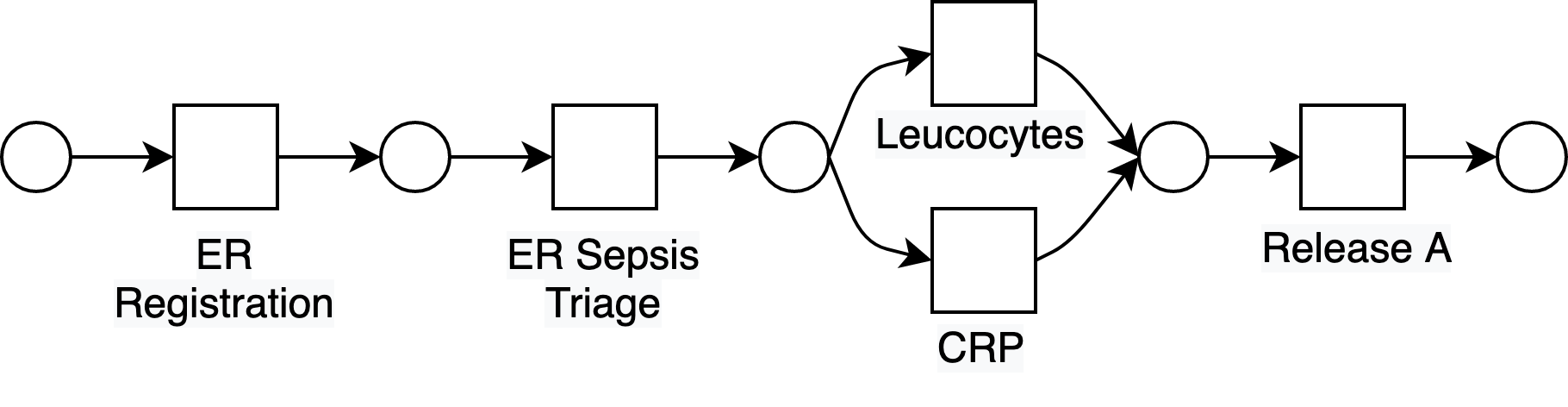}}
	\caption{Petri net for the~\emph{Sepsis} example}
	\label{fig-petrinet-sepsis}
\end{figure}

\paragraph{Petri nets} This representation is a specific kind of a bipartite graph~\cite{reisig2012petri}. There are two types of nodes: places and transitions. These can be connected by arcs representing the flow relations. Each place can hold several tokens, which collectively define the current state of the system. Changing the state by firing enabled transitions produces sequences. 
Petri nets are more expressive than directly-follows graphs, because they can explicit describe concurrency~\cite{van2019practitioner}. The Petri net in~\Cref{fig-petrinet-sepsis} shows that \emph{ER Sepsis Triage} can be followed by either \emph{Leucocytes} or \emph{CRP}. 

\paragraph{LTL (Linear Temporal Logic)} is the temporal logic with modalities that describe time. Event sequences can be described as a collection of rules by which the underlying behavior is constrained. One such specific modeling language that uses LTL for describing event sequence data is \emph{Declare}~\cite{DBLP:conf/cidm/MaggiMA11,DBLP:conf/simpda/PrescherCM14}. \Cref{table:declare-example} shows a set of rules showing that constrain event sequences. For example, the rule ChainSuccession(\emph{ER Registration}, \emph{ER Sepsis Triage}) means that that the first event type has to be directly followed by the second. 

\begin{table}[h] 
	\caption{Example of a Declare model for Sepsis example}
	\label{table:declare-example}
	\small
		\input{tables/example-sepsis-declare.tex}

\end{table}

\begin{figure}[h]
	\centering
	\resizebox{0.6\columnwidth}{!} {
		\includegraphics[width=\linewidth]{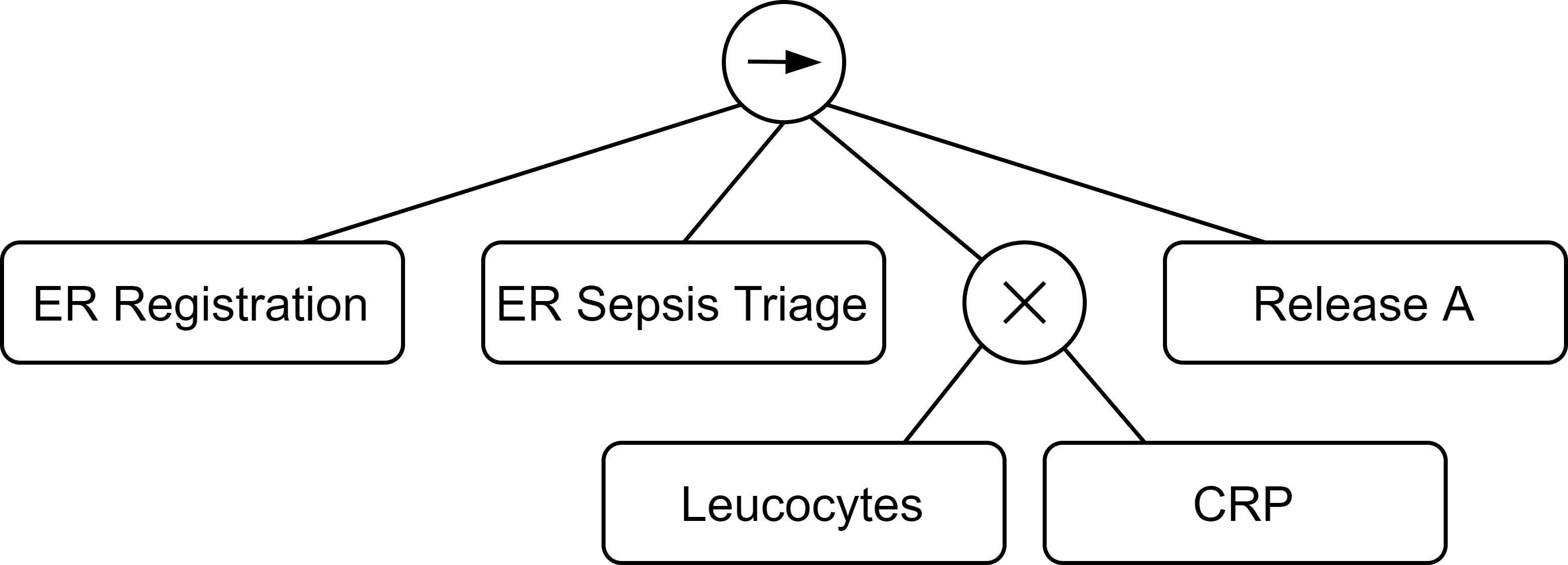}}
	\caption{Process tree for the~\emph{Sepsis} example}
	\label{fig-processtree-sepsis}
\end{figure}

\paragraph{Process Trees} are abstract hierarchical representations of a process model~\cite{arriagada2017techniques}. This tree represents events as leaf nodes, and control-flow operators as non-leaf nodes. There are a number of operators available such as exclusive choice $\times$, sequence $\rightarrow$, parallelism $\wedge$. A process tree for the Sepsis case is shown in~\Cref{fig-processtree-sepsis}. The sequence operator at the root of the tree indicates that the child nodes will be sequentially executed. The event sequence starts with \emph{ER Registration}, continues with \emph{ER Sepsis Triage}, which follows by either \emph{Leucocytes} or \emph{CRP}. The sequence finishes by the final event~\emph{Release A}.

%
%

\subsection{Visualization of Event Sequence Data}
\label{subsec:established-vis-rep}
\begin{figure}
	\centering
	\includegraphics[width=\linewidth]{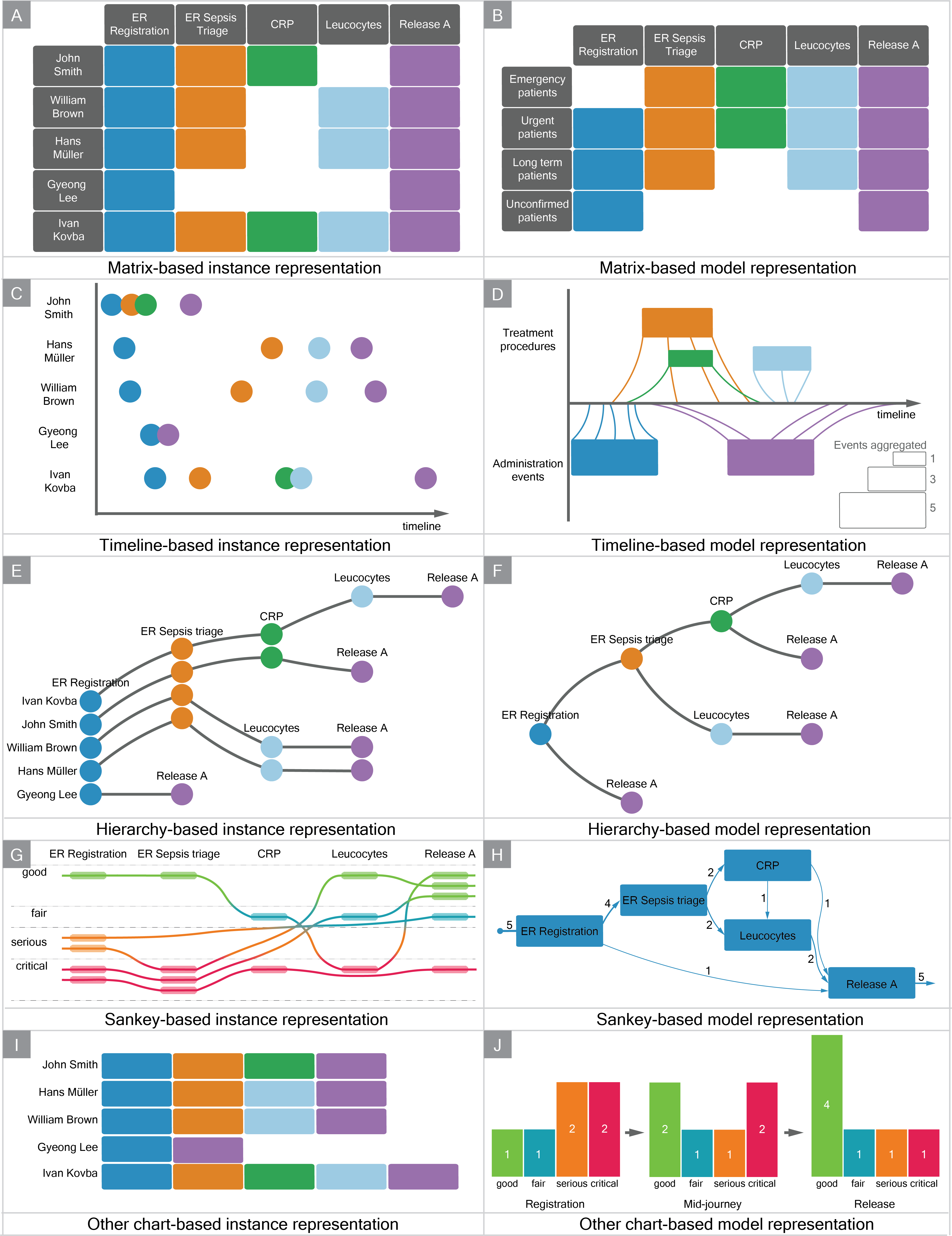}
	\caption{Example of representation using Sepsis use-case}
	\label{fig-representations}
\end{figure}

\noindent
There are various archetypes of how event sequence representations can be visualized. Here, we use the term visualization to refer to an arrangement of visual elements on a canvas.
In this section, we explain different visualization types that are used to support the analysis of event sequence data. We use the taxonomy presented in~\cite{DBLP:journals/corr/abs-2006-14291} to differentiate five main categories. We describe these categories in the order from most specific to least specific. In \Cref{fig-representations}, we exemplify each of these five categories for instance-based and model-based representations using the Sepsis example.


\subsubsection{Matrix Visualization}
\noindent
A matrix visualization is the most specific visualization of event sequences. 
The key feature of this visualization is that event sequences or aggregate values from event sequences are mapped to rows and columns of a matrix. Relationships between two orthogonal categories are shown as various glyphs at the matrix cell where both categories intersect. \Cref{fig-representations}.A shows an instance-based matrix visualization inspired by the design presented in~\cite{DBLP:journals/is/BoseA12}. Each row of this matrix represents one sequence from the Sepsis treatment process. The columns correspond to various events that appear in the event sequences. The model-based matrix visualization is inspired by the design of~\cite{DBLP:journals/tvcg/ChenAAABKNTT20} shown in~\Cref{fig-representations}.B. Here, the rows of the matrix correspond to clusters of patients based on event sequences and columns represent the events that are present in those categories.



\subsubsection{Timeline Visualization} 
\noindent
Timeline visualizations are visual representations of event sequences that use a time axis to align the elements. This visualization is usually used to convey the evolution of events over time, the evolution of aggregate values, or the order of events in event sequences. The example visualization shown in~\Cref{fig-representations}.C is based on the dotted chart design~\cite{DBLP:journals/is/BoseA12}. It uses circles to position each individual event on a horizontal timeline for each patient of the Sepsis use-case. The model-based representation of ToPIN~\cite{DBLP:journals/cgf/SungHSCLW17} is depicted in~\Cref{fig-representations}.D. It shows the events associated with different activities as they are executed in event sequences. In this way, it conveys information of the general order, number of events, and about groups of events (events are shown below or above the timeline).


\subsubsection{Hierarchy Visualization}
\noindent
Hierarchy visualizations are graph-based event sequence visualizations that use a tree topology. These visualizations convey hierarchical dependence and proximity of events in event sequences. Such visualizations can be generated using clustering algorithms.
\Cref{fig-representations}.E shows the design of individual event sequences with an order of events. The same activities are grouped together to convey the hierarchical relationships in the data. The visualization of~\cite{DBLP:journals/tvcg/VrotsouN19,DBLP:journals/cgf/LiuKDGHW17} is shown in~\Cref{fig-representations}.F. It shows the hierarchy-based model representation of event sequences, conveying information about general patterns in this data set.


\subsubsection{Sankey Visualization}
\noindent
Sankey visualizations are inspired by flow charts, showing events as bars and transitions as lines connecting them. Lines are proportional to the number of transitions in the event data. This visualizations type also includes node-link diagrams and directly-follows graph~\cite{DBLP:conf/icpm/LeemansPW19}. 
The visualization depicted in~\Cref{fig-representations}.G shows a sankey diagram for an instance-based representation of event sequences. Each event sequence, e.g. a sepsis patient, is visible as a line going through the bars of each event. The shown variant is based on the sequence braiding visualization ~\cite{DBLP:journals/tvcg/BartolomeoZSD21}. The model-based representation described in~\cite{DBLP:journals/is/BoltLA18} is depicted in~\Cref{fig-representations}.H. It shows a directly-follows graph for a sepsis case. This visualization arranges the data as a graph, showing the aggregated number of transitions between events.



\subsubsection{Other Chart Visualizations}
\noindent
Some visualizations are more basic or combinations of other archetypes.
We put all these other visualizations into this category. These typically include visualizations that combine simple visualization types such as bar charts, pie charts or box plots with means to convey some sequential information of the event sequence data. An instance representation inspired by~\cite{DBLP:journals/tvcg/UngerDSL18} is shown in~\Cref{fig-representations}.I. It uses a bar chart to represent individual sequences in the visualization. An example of model-based visualization is shown in~\Cref{fig-representations}.J. Here, the sequence of charts are used to convey information extracted from the raw event sequences. In this case, the progression of the patients' condition through the hospital process is displayed with bar charts. These charts are commonly used as additional views to the main visualization.

%


\subsection{Event Sequence Visualization Framework (ESeVis)}
\label{subsec:2x5}
\noindent
We have identified and discussed two orthogonal dimensions for describing event sequence data: instance-based or model-based representation as the conceptual axis and five different visualization types as the visual axis. Our Event Sequence Visualization Framework (ESeVis) builds on these two axes. Table~\ref{table:2x5} describes the $2\times 5$ categories of our framework. In the following, we will use ESeVis for categorizing contributions from the research fields of information visualization and process mining, in turn.

\begin{table} 
	\caption{ESeVis Framework}
	\label{table:2x5}
	\small
		\input{tables/2x5.tex}

\end{table}

%% file: tables/example-sepsis.tex
\rowcolors{2}{white}{gray!12.5}
\begin{tabular}{p{0.2\linewidth} p{0.74\linewidth}}
	\toprule
	\textbf{Steps}   & \textbf{Description} \\ \midrule
	ER Registration & Register a patient \\
	ER Triage  & Fill a general document to assign degree of an illness and decide the order of treatment (Triage document) \\ 
	ER Sepsis Triage& Fill an additional triage document if a patient is suspected to have sepsis \\ 
	CRP & Activity related to measurement of C-Reactive Protein (CRP) \\ 
	LacticAcid & Activity related to measurement of LacticAcid \\ 
	Leucocytes & Activity related to measurement of Leucocytes \\ 
	Release C & Patient discharge \\ \bottomrule
\end{tabular} 
%
%

%% file: tables/example-sepsis-declare.tex
\rowcolors{2}{white}{gray!12.5}
\begin{tabular}{p{0.46\linewidth} p{0.48\linewidth} }
	\toprule
	\textbf{Rule} & \textbf{Explanation}   \\ \midrule
	ChainSuccession(\emph{ER Registration}, \emph{ER Sepsis Triage})   & 
					If \emph{ER Registration} occurs then \emph{ER Sepsis Triage} occurs immediately afterwards
	 			     \\
	ChainPrecedence(\emph{ER Sepsis Triage}, \emph{Leucocytes})      &      
					If \emph{Leucocytes} occurs then \emph{ER Sepsis Triage} occurs immediately beforehand
					\\
	AtMostOne(\emph{ER Sepsis Triage})  & 
					If \emph{ER Sepsis Triage} occurs, then it occurs at most once
					\\
	NotSuccession(\emph{Release A}, \emph{CRP}) &
					\emph{Release A} occurs if and only if \emph{CRP} does not occur afterwards
					\\
	... & 
			        \\ \bottomrule
\end{tabular} 
%
%

%% file: tables/2x5.tex
\rowcolors{2}{white}{gray!12.5}
\begin{tabular}{r p{0.35\linewidth} p{0.35\linewidth} }
	\toprule
	\cellcolor{white} & \textbf{Instance representation} & \textbf{Model representation}   \\ \midrule
	Matrix-based    & 
					Event sequences and their features are mapped to the rows and columns of the matrix. Each matrix cell shows some values such as the presence of a particular event in the sequence.
					&
					Aggregated values coming from events, their additional values, or whole event sequences are mapped to rows and columns of the matrix. Each cell then shows some characteristics of these aggregations.
					\\
	Timeline-based  &
					Each event sequence is mapped to the visual elements on the visualization with a time axis.
					&
					The event sequences are aggregated before being mapped to the visual elements on the visualization with a time axis.
					\\
	Hierarchy-based & 
					The event sequences are mapped to the visualization that uses the color and a layout to show the hierarchical dependencies in the data.
					& 
					Event sequences are hierarchically clustered and mapped to the visualization elements that convey and highlight a hierarchical relationship present in event sequences. The event sequences are usually clustered with prefix-based clustering algorithms.
					\\

	Sankey-based    &      
					Each event sequence from a collection of event sequences is directly mapped to the visual elements of a Sankey-based visualization.
					&
					Sankey-based visualization displays event sequences after aggregations over the initial event sequences.
					\\
	Chart-based     & 
					Chart-based visualizations or their elements are used to map event sequence data to the visual elements.
					&
					Aggregate statistics derived from an event sequence are visualized with chart-based visualization or a set of chart-based visualizations.  
	\\
 \bottomrule
\end{tabular} 
%
%

%% file: sections/research-design.tex
\noindent
Research on visualizations of event sequence data have been published in major outlets of process mining and information visualization research.
In this section, we describe how we review these streams of literature. We follow guidelines as outlined in~\cite{kitchenham2004procedures,cooper2019handbook} and exemplified in~\cite{leitner2014systematic}. 
First, we explicate our overall review objectives. Then, we define our search and selection procedures. Finally, we report our classification procedures.

\subsection{Research objectives}
\noindent
The goal of this survey is to analyze the visualizations for event sequence data from two notable streams of research: information visualization and process mining. While both of these fields share commonalities in terms of data types, they have developed largely disconnected from each other. Our systematic literature review aims to categorize contributions from these two research fields in order to highlight commonalities and potential for mutual inspiration. We approach these objectives by focusing on the following research question (Q):

\begin{quote}\emph{Q1: Which type of event sequence visualizations have been proposed by which stream of research?}\end{quote}

We refine this research question into the following sub-questions:
\begin{itemize}
	\item  Q1.1: How can event sequence visualizations be classified?
	\item  Q1.2: How has research on event sequence visualizations developed over time?
	\item  Q1.3: What are the main difference between two research streams?
\end{itemize}


\subsection{Literature Search}
\noindent
We focused on the major process mining and information visualization journals for our systematic literature review. We consider the recent period between 2000-2020 for the search.\footnote{Note that the choice for this period is driven by the assumption that relevant visualizations should be in publications over such a span of time. Both fields look back at a more than 100 year old history~\cite{mendling-nordsieck,DBLP:series/hci/AignerMST11}.}
We selected three major research journals in the information visualization field that publish articles about event sequence visualizations. These journals are \acrfull{tvcg}, \acrfull{iv}, and \acrfull{cgf}. For process mining research, we selected the following three major journals: \acrfull{is}, \acrfull{tkde}, and \acrfull{dss}. 

Information visualization and process mining related publications use a different vocabulary to describe the works focusing on event sequence data. Therefore, we iteratively defined an initial keyword list, and extended it with the keywords extracted from relevant publications. 
For the visualization journals we searched for the following keywords: \emph{event-based data}, \emph{temporal event data}, \emph{timeline}, \emph{sankey}, \emph{process mining}. Looking through the results we identified additional keywords and added them to our search list: \emph{flow}, \emph{sequence mining}, \emph{action sequence}, \emph{Temporal visualization}, \emph{behaviour}, \emph{log data},  \emph{timenets}, \emph{Temporal Categorical}, \emph{sequential pattern mining}, \emph{event sequence}, \emph{Event sequences}, \emph{temporal event sequences}. 
For process mining focused journals we used the following search keywords: \emph{Process mining}, \emph{Event sequence}  and additional keywords \emph{Event sequences}.
For the selection of relevant papers, we used the Publish or Perish software~\cite{PublishorPerish} with the Google Scholar backend. We build our queries using the ISSN journal identifiers and mentioned keywords.

\subsection{Selection Procedure}
\begin{figure}[h]
	\centering
	\includegraphics[width=\linewidth]{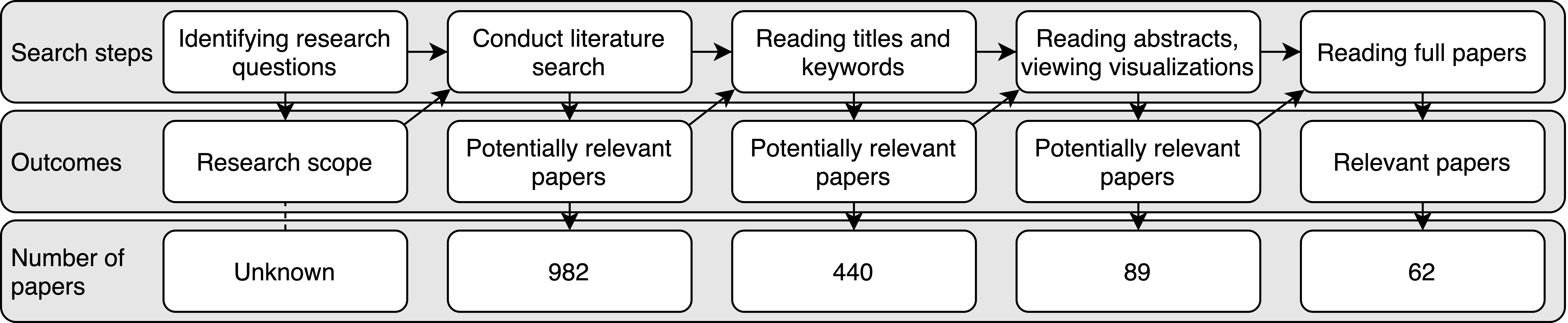}
	\caption{Literature search and selection process}
	\label{fig-selection-procedure}
\end{figure}

\noindent
The selection procedure has the aim to identify step by step relevant papers. The starting point are the papers identified by the search queries. \Cref{fig-selection-procedure} gives an overview of the procedure.

\begin{enumerate}
    \item We selected all the papers returned by the keyword search. 
    \item We analyzed the titles of each paper for being unrelated or potentially related based on our inclusion criterion. Our inclusion criterion considers research articles as relevant that propose new visualization techniques for event sequence data. 
    \item We analyzed abstracts and visualizations presented in papers. If a papers contains an abstract that discusses the visualization of event sequence data, or shows visualizations related to one of the categories described in~\Cref{subsec:established-vis-rep}, we consider it to be potentially relevant.
    \item We fully read the papers and identified the set of relevant papers. We documented the steps as summarized in~\Cref{table:selection}. 
    Overall, we identified 62 relevant papers.
\end{enumerate}


\begin{table} 
	\caption{Selection of papers}
	\label{table:selection}
	\centering
	\small
		\input{tables/selection.tex}

\end{table}

\subsection{Classification Procedure}
\noindent
The aim of the classification is to identify where which type of contributions have been made.
We analyzed each of the 62 included papers and their visualizations and categorized the contributions according to our ESeVis framework (see~\Cref{table:2x5}). We documented additional details about specifics of the identified visualizations. Often, research articles introduce visualization systems that contain one or more views to show a specific perspective on the event sequence data. Therefore, each contribution can belong to several categories in the 2x5 framework. Moreover, several visualizations introduce views that can be classified into more than one categories (e.g. the icicle plot in~\cite{DBLP:journals/cgf/LiuKDGHW17} is categorized into the timeline-based, model-based category and the hierarchy-based model representation as it adheres to definitions of both). Next, we present our findings.

%% file: tables/selection.tex
\rowcolors{2}{white}{gray!12.5}
\begin{tabular}{r r r r r}
	\toprule
	\cellcolor{white} & \textbf{Stage 1} & \textbf{Stage 2}  & \textbf{Stage 3} & \textbf{Stage 4} \\ \midrule
	\acrshort{tvcg}  & 339  & 180 & 56  & 37 \\
	\acrshort{iv}      & 66     & 34  & 6   & 1 \\
	\acrshort{cgf}    & 113   & 54  & 11  & 9 \\  \midrule
	Information Visualization            & 518   & 267  &  73 & 47 \\ \midrule[1.5pt]
	\acrshort{is}      & 195  & 108   & 9   & 9 \\
	\acrshort{tkde}  & 158   & 28   & 1    & 1 \\
	\acrshort{dss}   & 111   & 36    & 6    & 5 \\ \midrule
	Process Mining & 464 & 172  &  16 & 15 \\ \midrule[1.5pt]
	\textbf{Total}  & \textbf{982} & \textbf{440}  & \textbf{89} & \textbf{62} \\\bottomrule
\end{tabular} 
%
%

%% file: sections/results.tex
\noindent
In this section, we describe the results of our literature review. First, we will give an overview of the contributions. Then, we will discuss each of the five types of visualization and their sub-types in turn. 

\subsection{Overview of Contributions}
\noindent
The number of relevant papers is shown in~\Cref{table:selection}. 62 articles were selected from the six journals. We observe that TVCG appears to be the journal with the most articles on the visualization of event sequence data. More than half of the relevant papers, namely 37 of 62, are published in this journal. CGF and IS follow with 9 articles each. DSS has five articles, both IV and DSS one.

\begin{figure}[h]
	\centering
	\includegraphics[width=\linewidth]{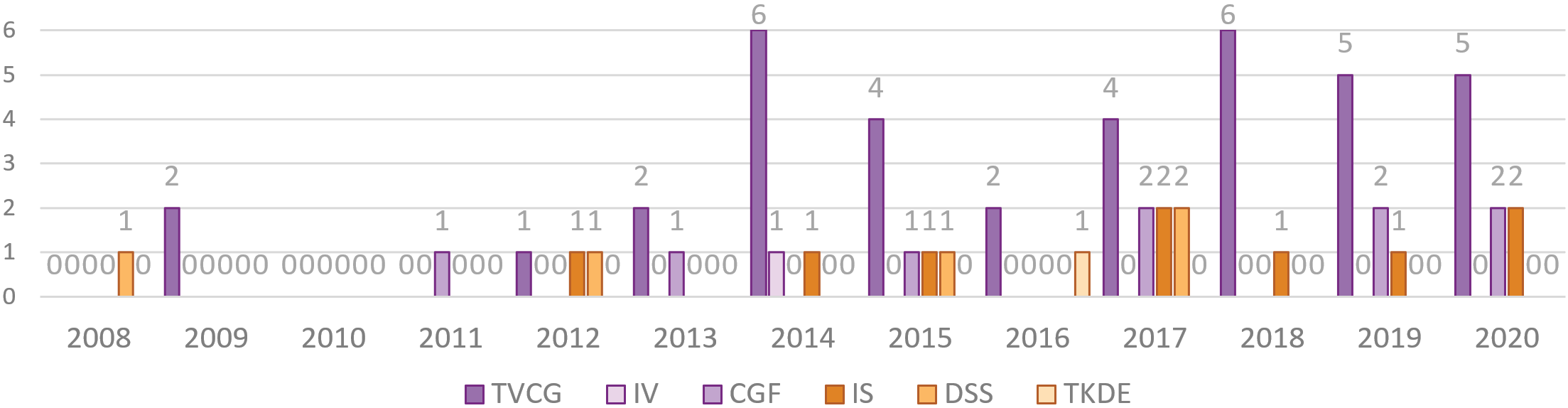}
	\caption{Publications by journal}
	\label{fig-papers-over-journal}
\end{figure}

\begin{figure}[h]
	\centering
	\includegraphics[width=\linewidth]{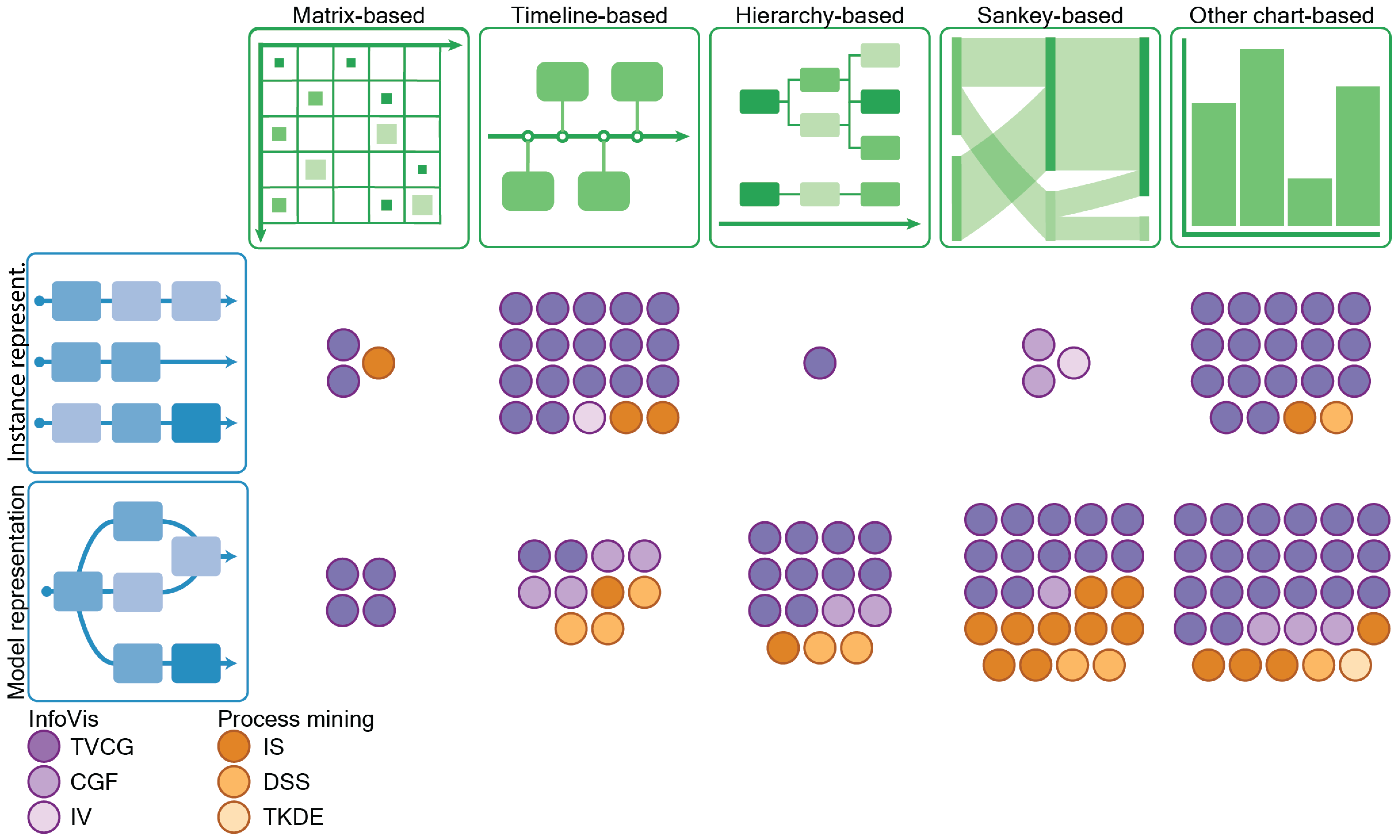}
	\caption{Publications with contributions in the different categories}
	\label{fig-2x5-icons-stats3}
\end{figure}

Research on the visualization of event sequence data has been developing in the twenty years since 2000. The distribution over years is shown for each journal in~\Cref{fig-papers-over-journal}. We observe that there are a few articles between 2008 and 2013. Only 2014 sees a drastic increase of published contributions. Since then, almost 10 articles are published each year, most of them in TVCG. \Cref{fig-2x5-icons-stats3} shows how the contributions of these articles are distributed over the different categories of visualizations for event sequence data. We will now discuss each category in detail. We use "IV" for information visualization and "PM" for process mining together with the reference number of the paper to indicate to which category a reference belongs.

\subsection{Matrix-based Visualizations}
\noindent
Matrix-based visualizations use a matrix layout to display characteristics of event sequence data. This visual representation has two dimensions, where the rows correspond to one characteristic such as unique event sequence identifiers, and columns correspond to another dimension of data, such as events in a sequence, or transitions between events. The summary of contributions that use a matrix-based visualizations is presented in~\Cref{table:matrix}. Next, we separately discuss {\matrixA} Visualizations and then {\matrixB} Visualizations.

\begin{table} 
	\caption{Articles covering matrix-based visualizations}
	\label{table:matrix}
	\centering
	\tiny
	\setlength\tabcolsep{1.5pt} 
		\input{tables/4-matrix}
\end{table}

\subsubsection{\matrixA Visualizations} 

\begin{wrapfigure}[7]{l}{0.25\textwidth}
	\centering
	\includegraphics[width=0.25\textwidth]{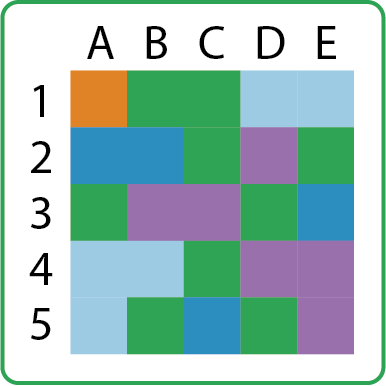}
	\caption{{\matrixA} visualization}
	\label{fig-matrix}
\end{wrapfigure}

\noindent
In {\matrixA} visualizations, the matrix cells are used to represent elementary information that can be described as a categorical values (see~\Cref{fig-matrix}). The cells are usually distinguished with color, which represent the type of the cell. 

Bose et al.~\cite[PM]{DBLP:journals/is/BoseA12} use the matrix layout to display event sequence data at the sequence representation level. They arrange event sequences in a matrix view as follows: each line represents one event sequence and each column represents an event type. This visualization supports the analysis and complements a corresponding algorithm that minimizes the number of vertical gaps in the matrix for aligning similar event sequences.

Other works use {\matrixA} visualizations to show aggregated characteristics of event sequence data~\cite[IV]{DBLP:journals/tvcg/ChenAAABKNTT20,Kwon2020DPVisVA} by conveying categorical information with color hue. Chen et al.~\cite[IV]{DBLP:journals/tvcg/ChenAAABKNTT20} use topic modeling algorithms to aggregate event sequences into topics, and generate visualizations showing how these topics map to events in a matrix form. Kwon et al.~\cite[IV]{Kwon2020DPVisVA} use a matrix view to show medical data using hidden markov models (HMM). They lay out the discovered states of these models as columns and multiple event sequence attributes as rows. 

\subsubsection{\matrixB Visualizations} 
\begin{wrapfigure}{l}{0.25\textwidth}
	\centering
	\includegraphics[width=0.25\textwidth]{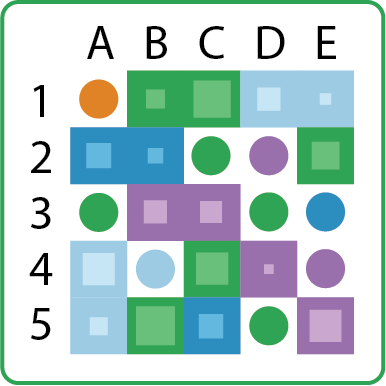}
	\caption{{\matrixB} visualization}
	\label{fig-matrix-extended}
\end{wrapfigure}

\noindent
Some matrix visualizations are more complex. They use the cells in order to convey additional quantitative and qualitative information through a special glyph designs. This design is used to convey more information than just simple categorical values (~\Cref{fig-matrix-extended}). 

Loorak et al.~\cite[IV]{DBLP:journals/tvcg/LoorakPKHC16} represent the medical data of stroke patients. Instead of showing the raw event sequences only, they use a matrix visualization to convey information about additional parameters attached to each sequence. In this way, the matrix cells show quantities as bar charts, categorical information with color, and shapes.

Some {\matrixB} visualizations convey aggregated information about event sequences. Jin et al.~\cite[IV]{DBLP:journals/corr/abs-2009-00219} use a matrix view to present a causality analysis for event sequence data. They use the rows of the matrix to represent causes and the columns to show effects. Each cell of a matrix is split into an outer and an inner region to represent two casual relation groups. Color saturation is used to represent the strength of the corresponding relations. Another usage of a matrix view is presented by Xie et al.~\cite[IV]{DBLP:journals/corr/abs-2009-02464}. They use a matrix with a custom glyph design to represent passes in football.

\subsection{Timeline-based Visualizations}
\noindent
A timeline-based visualization is a visual representation of event sequence data that has a fixed time axis (usually the x-axis) to arrange visual elements that represent events or derived properties in temporal order. Next, we describe several categories of timeline-based visualizations. For a summary of papers in this category refer to~\Cref{table:timeline}.

\begin{table} 
	\caption{Papers presenting timeline-based visualizations}
	\label{table:timeline}
	\centering
	\tiny
	\setlength\tabcolsep{1.5pt} 
	\input{tables/4-timeline}
\end{table}

\subsubsection{\timelineA} 

\begin{wrapfigure}[7]{l}{0.25\textwidth}
	\centering
	\includegraphics[width=0.25\textwidth]{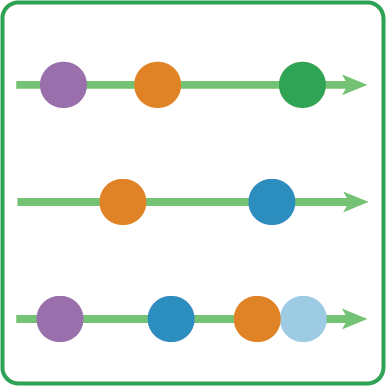}
	\caption{{\timelineA} visualization}
	\label{fig-timeline-fixed}
\end{wrapfigure}

\noindent
The first category is called {\timelineA} visualizations and is depicted in~\Cref{fig-timeline-fixed}. Visualizations in this category present event sequences as elements that are aligned along the timeline. 

Bose et al.~\cite[PM]{DBLP:journals/is/BoseA12} use a representation of event sequences, in which each event has one timestamp (beginning or the end of event execution). This visualization is called dotted chart. An event sequences is shown as a sequence of circular glyphs colored by event type. Each sequence is arranged as a row. Similar visualization are used in several visual systems~\cite[IV]{DBLP:journals/tvcg/ChenXR18,Kwon2020DPVisVA,DBLP:journals/cgf/LeiteGMGK20,DBLP:journals/cgf/DortmontEW19}, with different glyph shapes including triangles~\cite[IV]{DBLP:journals/cgf/LeiteGMGK20,DBLP:journals/tvcg/MonroeLLPS13} and rhombus~\cite[IV]{DBLP:journals/cgf/DortmontEW19}. Other works also use {\timelineA} visualizations~\cite[IV]{DBLP:journals/tvcg/FuldaBM16,DBLP:journals/tvcg/MonroeLLPS13}, but combine them with a {\timelineB} visualization that is described as a next category. 

Some works use a {\timelineA} visualizations to represent \emph{aggregated} event sequences. The visualization system by~\cite[IV]{DBLP:journals/cgf/LeiteGMGK20} shows the colored vertical rectangles on the timeline to visualize all events and when they happened on the timeline. de Leoni et al.~\cite[PM]{DBLP:journals/dss/LeoniAAH12} use a timeline view with circular glyps to show the events that are planned to be performed in the future. Grouping of several circles creates a bigger circle that is used as a pie chart to represent if some of the contained events were already executed.

\subsubsection{{\timelineB}}

\begin{wrapfigure}{l}{0.25\textwidth}
	\centering
	\includegraphics[width=0.25\textwidth]{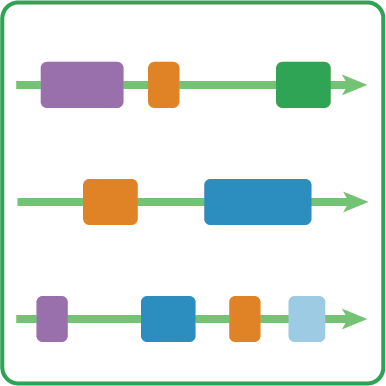}
	\caption{{\timelineB} visualization}
	\label{fig-timeline-flexible}
\end{wrapfigure}
\noindent
In this event sequence visualization type, each event is mapped to a timeline with duration information encoded as a length of a glyph (see~\Cref{fig-timeline-flexible}).

The authors of~\cite[IV]{DBLP:journals/cgf/HanRDAS15,DBLP:journals/tvcg/NguyenTAATZ19,DBLP:journals/tvcg/VrotsouN19,DBLP:journals/tvcg/ZengWWWLEQ20} and~\cite[PM]{DBLP:journals/is/LowAHWW17} use a horizontal timeline to arrange the rectangular visual elements colored according to their event type. In this way, they convey the time from the beginning to the end of the event, often to capture the duration of an activity execution. Another group of works~\cite[IV]{DBLP:journals/tvcg/VrotsouJC09,DBLP:journals/ivs/VrotsouYC14} encodes events as a color-coded rectangle and aligns them to the vertical timeline. Rosenthal et al.~\cite[IV]{DBLP:journals/cgf/RosenthalPMO13} describe a visualization system that shows problems and their severity with a color of a standard sized boxes, while the lead time (completion time) is shown as a line extending past the box instead of the enlarged box size.
Other works, instead of showing individual events from event sequence data, focus on conveying some information derived from event sequences on the timeline. Richter et al.~\cite[PM]{DBLP:journals/is/RichterS19} use a duration timeline-based visualization to show when particular transitions between events were performed. 

The two categories discussed so far are often considered to be pure event-based visualizations. They convey information about the time of when an event happened as one dimension of the visualization of the event sequence. The following two categories deal with more complex visualizations, in which additional information is integrated.

\subsubsection{{\timelineC}}

\begin{wrapfigure}[7]{l}{0.25\textwidth}
	\centering
	\includegraphics[width=0.25\textwidth]{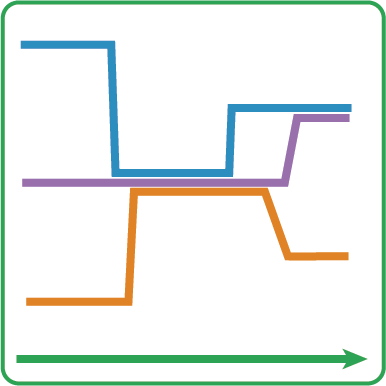}
	\caption{{\timelineC} visualization}
	\label{fig-timeline-converging}
\end{wrapfigure}

\noindent
This type of timeline visualization illustrated in~\Cref{fig-timeline-converging} displays event sequences as converging and diverging lines. Each of these lines corresponds to an event sequence. The lines that lay close together usually correspond to some common characteristic such as location proximity, same state, or joint resource executing these events. 

This category has only been discussed in the field of information visualization. Some works build on Marey graphs to represent event sequences. Xu et al.~\cite[IV]{DBLP:journals/tvcg/XuMR017} use a Marey graph to represent the assembly line performance at a factory, showing progression of events as vertical lines that follow a sequence of  predefined events along the graph. This visualization uses color to highlight those parts of the lines that represent delays. Another Marey graph based visualization is shown in~\cite[IV]{Kwon2020DPVisVA}, using color to highlight the event types for the use case of disease progression pathways.

Baumgartl et al.~\cite[IV]{DBLP:journals/corr/abs-2008-09552} use a vertical {\timelineC} visualization in a system for the analysis of pathogen transmission in hospitals. The horizontal interactive timeline contains the lines representing patients and their health status. The lines that converge show the patient contacts. The visualization by Liu et al. presented in~\cite[IV]{DBLP:journals/tvcg/LiuWWLL13} uses lines to represent the characters of a story and events that involve them as converging lines of several lines. Another {\timelineC} visualization by Reda et al.~\cite[IV]{DBLP:journals/cgf/RedaTJLB11} shows the evolution of how social groups evolve with time. The lines that stretch horizontally represent the group membership, while diverging and converging of the lines convey the change of the group membership. 

\subsubsection{{\timelineD}}

\begin{wrapfigure}{l}{0.25\textwidth}
	\centering
	\includegraphics[width=0.25\textwidth]{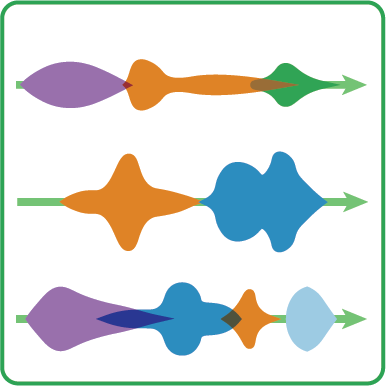}
	\caption{{\timelineD} visualization}
	\label{fig-timeline-evolution}
\end{wrapfigure}

\noindent
The {\timelineD} visualization (\Cref{fig-timeline-evolution}) conveys event sequence information as a density chars arranged along the timeline. Event types are often distinguished by color and the area size is used to show additional characteristics.

Wu et al.~\cite[IV]{DBLP:journals/tvcg/WuLYLW14} introduce the Opinionflow visualization that shows the opinion diffusion on social media as a sankey-based-density map visualization. In this visualization, where the event sequences are aggregated in several horizontal timelines. The size of the area shows the number of opinion expressions on social media and the color shows the type of opinion. Sung et al.~\cite[IV]{DBLP:journals/cgf/SungHSCLW17} present a {\timelineD} view as part of their visualization system. This view is called a theme river and it summarizes a number of events as a color-highlighted area displayed over a horizontal timeline. A similar view is presented as part of the ChronoCorrelator visualization system~\cite[IV]{DBLP:journals/cgf/DortmontEW19}. It includes a density plot showing the number of all events that happen as an area arranged with a horizontal timeline.

The papers in this category focus on displaying the features derived from event sequence data in the timeline-based graphs. Suriadi et al.~\cite[PM]{DBLP:journals/dss/SuriadiOAH15,DBLP:journals/dss/SuriadiWXAH17} use line graphs to show how different event sequence parameters change with time. These graphs allow for inspection of the time-related changes in the event sequences.

\subsubsection{\timelineE}
\noindent
In this section we discuss visualization systems that combine several visualization categories in their designs.
 
The visualizations presented in~\cite[IV]{DBLP:journals/tvcg/MonroeLLPS13,DBLP:journals/tvcg/FuldaBM16} show a combination of {\timelineA} and {\timelineB} visualization systems. They use triangular~\cite[IV]{DBLP:journals/tvcg/MonroeLLPS13} and circular~\cite[IV]{DBLP:journals/tvcg/FuldaBM16} glyphs to represent the individual events, and range-plot visual elements to convey the information about events with a duration. 
The ToPIN visualization~\cite[IV]{DBLP:journals/cgf/SungHSCLW17} captures the online learners behavior with a ToPIN timeline-based visualization as a combination of {\timelineA} and {\timelineB}. In this visualization different events from event sequences are grouped in the rectangles that convey the average time range of when the events grouped to this rectangle happened. Additionally, these rectangles connected with a timeline axis with curved lines show the exact timestamps of individual events that they contain.

The LiveGantt visualization~\cite[IV]{DBLP:journals/tvcg/JoHPKS14} combines {\timelineB} and {\timelineD} designs. The visual elements are placed on the horizontal timeline with the length representing the duration of the event, and the height representing the number of events of the same type that occurred at a given time.
The Coreflow visualization~\cite[IV]{DBLP:journals/cgf/LiuKDGHW17} presents the main visualization that shows the tree-like patterns discovered from event sequences. Coreflow is presented in two ways, as a icicle plot and as a node-link visualization showing branching patterns of event sequence data. Both visualizations also preserve aggregated timestamps for each event. While this visualization is also showing a hierarchy-based representation, it is also a {\timelineB} and {\timelineD} visualization.

\subsection{Hierarchy-based}
\noindent
Hierarchy-based visualizations convey hierarchical relationships found in event sequence data. These visualizations group sequences by their prefix similarity. A summary of contributions that design visualization systems using this visual representation is shown in~\Cref{table:hierarchy}.

\begin{table} 
	\caption{Selection of papers for timeline-based visualizations}
	\label{table:hierarchy}
	\centering
	\tiny
	\setlength\tabcolsep{1.5pt} 
		\input{tables/4-hierarchy}

\end{table}

\subsubsection{\hierarchyA}
\begin{wrapfigure}{l}{0.25\textwidth}
	\centering
	\includegraphics[width=0.25\textwidth]{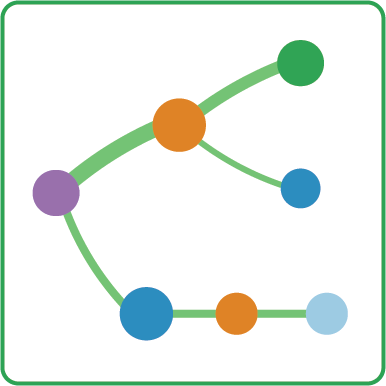}
	\caption{{\hierarchyA} visualization}
	\label{fig-hierarchy-node}
\end{wrapfigure}

\noindent
This type of visualizations (\Cref{fig-hierarchy-node}) captures information about event sequences as a node-link diagram, where nodes usually represent events and the links represent transition between events. 

In many visualization designs an elementary node-link diagrams is used, e.g. in~\cite[IV]{DBLP:journals/cgf/LeiteGMGK20,DBLP:journals/tvcg/LawLMB19} and~\cite[PM]{DBLP:journals/is/BoseA12,DBLP:journals/dss/SongA08,DBLP:journals/dss/LeoniAAH12}. Law et al.~\cite[IV]{DBLP:journals/tvcg/LawLMB19} and Bose et al.~\cite[PM]{DBLP:journals/is/BoseA12} use color to distinguish different event types. Also Leite et al.~\cite[IV]{DBLP:journals/cgf/LeiteGMGK20} and de Leoni et al.~\cite[PM]{DBLP:journals/dss/SongA08,DBLP:journals/dss/LeoniAAH12} use various glyph designs for event types additionally to color.

A second group of {\hierarchyA} visualization designs consider the node size, link size or both to show additional dimension in the data. For instance, Vrotsou et al.~\cite[IV]{DBLP:journals/tvcg/VrotsouN19,DBLP:journals/cgf/LiuKDGHW17} use the size of a link in node-link diagram to represent the frequency of the corresponding transition between two events. The visualization design by Krause et al.~\cite[IV]{DBLP:journals/tvcg/KrausePS16} takes the size of the links, size of nodes, and the color of nodes to convey the information about the frequencies and additional characteristics of a cohort of event sequence data.

While many hierarchy-based visualizations aggregate the event sequence data, the work by Vrotsou et al.~\cite[IV]{DBLP:journals/tvcg/VrotsouJC09} defines a node-link based visualization that only groups sequences containing the same sub-sequence of activities using user defined queries. This visualization shows hierarchically grouped sequences that are still individually identifiable. 

\subsubsection{\hierarchyB}

\begin{wrapfigure}{l}{0.25\textwidth}
	\centering
	\includegraphics[width=0.25\textwidth]{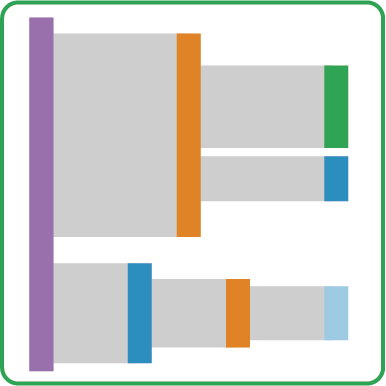}
	\caption{{\hierarchyB} visualization}
	\label{fig-hierarchy-flow}
\end{wrapfigure} 

\noindent
{\hierarchyB} visualizations aggregate event sequences in the sankey based visualization that shows the hierarchical dependencies in the event sequence data (see~\Cref{fig-hierarchy-flow}). These visualizations use the horizontal or vertical layouts with sequentially ordered elements, in which the width of a transition shows its frequency. 

Liu et al.~\cite[IV]{DBLP:journals/cgf/LiuKDGHW17} show Coreflow and its icicle plot as {\hierarchyB} visualization. This vertical design shows event sequences grouped by prefixes starting from the top of the visualization where similar prefixes are laid out proceeding downward to more specific events. The width of the events show the number of transition. Similar design, but laid out horizontally is used in the Outflow visualization~\cite[IV]{DBLP:journals/tvcg/WongsuphasawatG12}. In this visualization, the patterns are grouped and colored by an outcome by the scale green to red for a positive or a negative outcome. The EventThread system~\cite[IV]{DBLP:journals/tvcg/GuoXZGZC18} extends the EventFlow visualization and its icicle plot design with color to highlight the different event types. Jang et al.~\cite[IV]{DBLP:journals/tvcg/JangER16} present the MotionFlow system that uses glyphs to identify events, and the transition width and color to show separate aggregated event sequences.

Several special cases of {\hierarchyB} visualizations are described in~\cite[IV]{DBLP:journals/tvcg/GuoXZGZC18,DBLP:journals/tvcg/LiuWDHWW17,DBLP:journals/tvcg/LawLMB19,DBLP:journals/tvcg/GuoJGDZC19}. Guo et al.~\cite[IV]{DBLP:journals/tvcg/GuoXZGZC18,DBLP:journals/tvcg/GuoJGDZC19} present a system for the analysis of event sequence data that splits event sequences into stages. Their systems use nodes of a horizontally laid out graph to represent information about stages of event sequences. These nodes are connected with semi-transparent lines that use width to indicate the frequency of a certain path. Law et al.~\cite[IV]{DBLP:journals/tvcg/LawLMB19} show the MAQUI system in which the workspace view illustrates patterns as the vertical colored rectangles. Event transitions are shown with the gray horizontal rectangles using height to indicate the frequency of a transition. Finally, the system by Liu et al.~\cite{DBLP:journals/tvcg/LiuWDHWW17} shows the icicle view for patterns for different groups of users. Unlike the standard icicle plot, this visualization shows separate patterns for each group.

\subsubsection{\hierarchyC}

\begin{wrapfigure}[9]{l}{0.25\textwidth}
	\centering
	\includegraphics[width=0.25\textwidth]{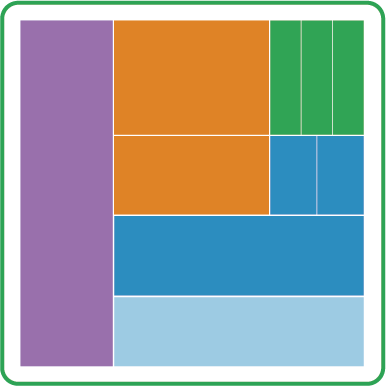}
	\caption{{\hierarchyC} visualization}
	\label{fig-hierarchy-treemap}
\end{wrapfigure} 

\noindent
This category includes visualizations of the event sequences that are aggregated into a treemap view (\Cref{fig-hierarchy-treemap}). This visualization uses nested figures such as rectangles to represent hierarchical dependencies in the data. 

The Motionflow system by Jang et al.~\cite[IV]{DBLP:journals/tvcg/JangER16} features a treemap view to complement a {\hierarchyB} view. It uses a nested rectangle layout, coloring the rectangles in the same way as the paths in the flow visual representations. This view then shows a hierarchical relationship of which patterns exists in the event sequence data.

\subsection{Sankey-based}
\noindent
Sankey-based visualizations show event sequence data as a graph of transitions between event types. Articles that employ sankey-based visualization designs are summarized in~\Cref{table:sankey}.

\begin{table} 
	\caption{Selection of papers for sankey-based visualizations}
	\label{table:sankey}
	\centering
	\tiny
	\setlength\tabcolsep{1.5pt} 
		\input{tables/4-sankey.tex}
\end{table}

\subsubsection{\sankeyA}
\begin{wrapfigure}{l}{0.25\textwidth}
	\centering
	\includegraphics[width=0.25\textwidth]{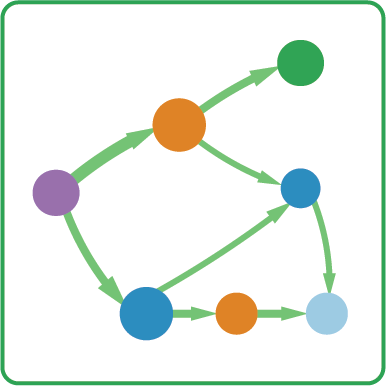}
	\caption{{\sankeyA} visualization}
	\label{fig-sankey-node}
\end{wrapfigure}

\noindent
The category of {\sankeyA} visualizations use a graph layout in which the events are represented as nodes and the transitions are shown as links (see~\Cref{fig-sankey-node}). A node-link visualization can convey additional information by the size of the node or the width of links such as frequency of occurrence of a specific type of transition.  

The authors of~\cite[PM]{DBLP:journals/is/BoseA12,DBLP:journals/is/WinterSR20} use node-link diagrams as an aggregated model derived from event sequence data. They annotate the transitions and nodes with frequencies. Song et al.~\cite[PM]{DBLP:journals/dss/SongA08} show the social network derived from event sequences as a node-link based representation. They use colors to distinguish event types. Baumgartl et al.~\cite{DBLP:journals/corr/abs-2008-09552} use a node-link representation to show the contact network view for the data of pathogen transmission between patients in the hospital. They use color to highlight the patient status and their contacts. The ViSeq visualization~\cite[IV]{DBLP:journals/tvcg/ChenYPCSPQ20} defines a node-link design that arranges the events vertically and the transitions as arcs.

\subsubsection{\sankeyB}
\begin{wrapfigure}{l}{0.25\textwidth}
	\centering
	\includegraphics[width=0.25\textwidth]{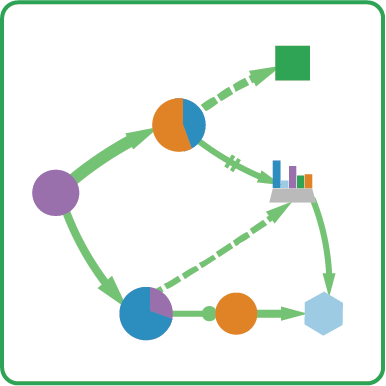}
	\caption{{\sankeyB} visualization}
	\label{fig-sankey-ex-node}
\end{wrapfigure}

\noindent
{\sankeyB} visualizations combine designs that convey information beyond the transitions between event types and their frequencies (see~\Cref{fig-sankey-ex-node}). These include characteristics such as the type of transition, the distribution of events as a single node, and nodes with different semantics (such as Petri nets~\cite[PM]{DBLP:journals/is/Munoz-GamaCA14}).

Some works distinguish \emph{transition types} in the node-link visualizations. Leoni et al.~\cite[PM]{DBLP:journals/is/LeoniMA15} show the node-link graph that describes the Declare rules applying to pairs of events. They add arrows, circles, and other glyphs on top of transitions to indicate the rule type as a transition. Bolt et al.~\cite[PM]{DBLP:journals/is/BoltLA18} use colors of transitions and dashed lines to represent differences between data and some predetermined rules. 

There are works that describe node designs that convey additional information. Leite et al.~\cite[IV]{DBLP:journals/cgf/LeiteGMGK20} uses the color and shape of the events to show the frequency and the type of the node. Jang et al.~\cite[IV]{DBLP:journals/tvcg/JangER16} define water-drop nodes in Motionflow that contain representations of the human motion patterns. The transitions in this node-link diagram show the change of a person's position. So-called Extended Compliance Rule Graphs by Knuplesch et al.~\cite[PM]{DBLP:journals/is/KnupleschRK17} visualize the compliance of an event sequences with the model. These graphs use different node shapes to represent various node types. Song et al.~\cite[PM]{DBLP:journals/dss/SongA08} adopt Petri nets and color the transitions with two-color schema (color of a circle and its border) to show performance measures.

Low et al.~\cite[PM]{DBLP:journals/is/LowAHWW17} use pie charts as nodes to show the extent of swap of resources that work on events. EmoCo~\cite[IV]{DBLP:journals/tvcg/ZengWWWLEQ20} define views that convey additional information about the aggregated events as colored pie charts. Wynn at al.~\cite[PM]{DBLP:journals/dss/WynnPXHBPA17} present Petri nets in the three-dimensional space including bar charts on the nodes to represent additional information, such as injury severity. Munoz-Gama et al.~\cite[PM]{DBLP:journals/is/Munoz-GamaCA14} also use Petri nets to support visual conformance checking, as much as van Dongen et al.~\cite[PM]{DBLP:journals/corr/abs-2011-11551} who convey additional information as horizontal bar charts in the nodes and various transition designs (solid lines, dashed lines, etc.).

\subsubsection{\sankeyC}

\begin{wrapfigure}{l}{0.25\textwidth}
	\centering
	\includegraphics[width=0.25\textwidth]{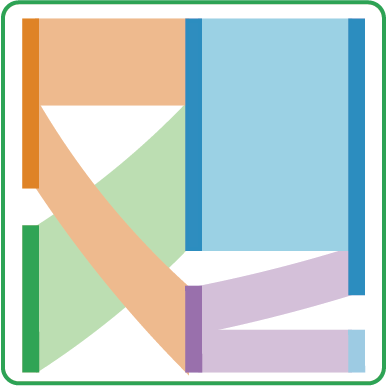}
	\caption{{\sankeyC} visualization}
	\label{fig-sankey-flow}
\end{wrapfigure} 

\noindent
In the {\sankeyC} visualizations the event sequences are aggregated showing the transitions of events in the event sequences (see~\Cref{fig-sankey-flow}). These visualizations are horizontally arranged from left to right according to the temporal order of events, and the width of transitions show the frequency of transitions.

Nguyen et al.~\cite[IV]{DBLP:journals/tvcg/NguyenBJKBGMB21} present a sankey-based flow visualization, in which each event type is associated with a histogram to show its variability across processes. Gotz et al.~\cite[IV]{DBLP:journals/tvcg/GotzZWSB20} use a sankey diagram to aggregate the flow of events representing them as colored rectangles and their size scaled to their aggregated frequency. 
Jin et al.~\cite[IV]{DBLP:journals/corr/abs-2009-00219} arrange cause-effect relations in a sankey diagram. The RoseRiver visualization system~\cite[IV]{DBLP:journals/tvcg/CuiLWW14} shows the change in groups of event sequences as a sankey-inspired diagram. For each step, color indicates the event sequence group and height frequency. The Loyaltracker~\cite[IV]{DBLP:journals/tvcg/ShiWLZQ14} visualizes the loyalty of search engine users. They use color hue to differentiate between three groups users: strictly loyal, medium loyal, and barely loyal. The flow of users represents those that join or leave a search engine at each point in time. The Opinion Flow system~\cite[IV]{DBLP:journals/tvcg/WuLYLW14} shows the information about the flow of interconnected events as a sankey-based-density map, where the color intensity and hue capture the state of the system and the separate flows indicate the frequency of a particular opinion discovered from raw event sequence. On demand, parts of the raw event sequence with the directly follows relations can be shown. The EmoCo visualization~\cite[IV]{DBLP:journals/tvcg/ZengWWWLEQ20} is a three-steps sankey diagram showing facial-emotion change along the sequence of particular words.  
	
The STBins visualization system~\cite[IV]{DBLP:journals/tvcg/QiBWWW20} is a sankey-based design for analyzing multi-thread execution log data. Event sequences are grouped in threads and visually encoded as rectangles containing circles that represent events. The transitions between circles within the thread and across threads encode the change in sequences.

Guo et al.~\cite[IV]{DBLP:journals/tvcg/GuoJGDZC19} use several chart-based visualizations (area charts, bar charts) in combination with a {\sankeyC} visualization to convey different characteristic of event aggregations.

\subsubsection{\sankeyD}

\begin{wrapfigure}[7]{l}{0.25\textwidth}
	\centering
	\includegraphics[width=0.25\textwidth]{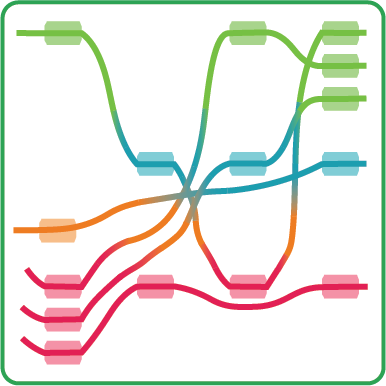}
	\caption{{\sankeyD} visualization}
	\label{fig-sankey-converging}
\end{wrapfigure} 

\noindent
{\sankeyD} visualizations have in common that they use horizontal lines that converge and diverge showing change of state of a sequence or a group of sequences (~\Cref{fig-sankey-converging}). 

Bartolomeo et al.~\cite[IV]{DBLP:journals/tvcg/BartolomeoZSD21} present a Sequence Braiding visualization system that uses vertically lines grouped into categories and highlighted with color. These lines show how the status diabetic patients change with events of food consumption. Chou et al.~\cite[IV]{DBLP:journals/cgf/ChouWM19} present a similar visualization design, grouping similar sequences with the same color. Vrotsou et al.~\cite[IV]{DBLP:journals/ivs/VrotsouYC14} define a related design, but instead of arranging the events vertically by one categorical measure, each step is represented based on quantitative measure. This visualization focuses on information derived from event sequence data, and not on the sequential order.

Agarwal et al.~\cite[IV]{DBLP:journals/cgf/AgarwalB20} present a {\sankeyD} design that represents group membership of event sequence changes over time. Line width indicates additional characteristics and bar chart display distribution information related to an event type.

\subsubsection{\sankeyE}

\begin{wrapfigure}[9]{l}{0.25\textwidth}
	\centering
	\includegraphics[width=0.25\textwidth]{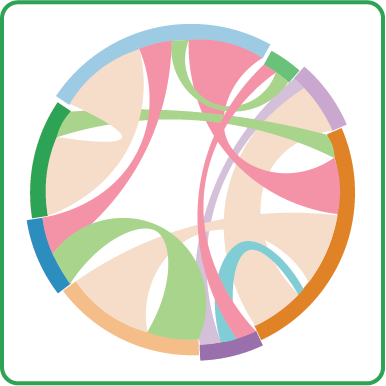}
	\caption{{\sankeyE} visualization}
	\label{fig-sankey-chord}
\end{wrapfigure} 

\noindent
The {\sankeyE} visualization captures events as circles with transitions as arcs (see~\Cref{fig-sankey-chord}). 

Chen et al.~\cite[IV]{DBLP:journals/tvcg/ChenYPCSPQ20} present a {\sankeyE} representation as part of their visualization system. It is used for showing aggregated transitions of event sequence data in different time windows (weeks). The width of each flow in this chord diagram shows the frequency of this transition. The design allows for better comparison between different event sequence groups.

\subsection{Other Chart-based}
\noindent
This group of event sequence data visualization designs use well-established visualization charts to display information derived from event sequence data. The examples of these charts are bar chart, histogram, pie chart, and sequence chart. Visualization systems often integrate a combination of several chart-based visualizations in one design. A summary of related contributions is shown in~\Cref{table:chart}.

\begin{table} 
	\caption{Selection of papers for chart-based visualizations}
	\label{table:chart}
	\centering
	\tiny
	\setlength\tabcolsep{1.5pt} 
		\input{tables/4-chart}

\end{table}

\subsubsection{Frequently used Chart Types}
\noindent
In this section we describe frequently used chart types for visualizing event sequences.

\paragraph{A. {\chartA}}

\begin{wrapfigure}{l}{0.25\textwidth}
	\centering
	\includegraphics[width=0.25\textwidth]{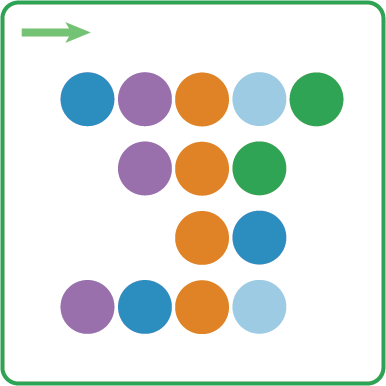}
	\caption{{\chartA} visualization}
	\label{fig-chart-A}
\end{wrapfigure}

In this category, events of sequences are mapped to visual elements such as different glyphs or shapes. Color is often used to show the event type (see~\Cref{fig-chart-A}). Visualizations in this category are arranged horizontally or vertically, such that the order of events is easily understood. In two-dimensional layouts, lines are used to connect glyphs. {\chartA} visualizations are similar to {\timelineA}, but without the requirement to show elements arranged along the timeline. 

The first group of sequence chart-based works focus on providing an overview of event sequences. They visualize events as sequence of circles~\cite[IV]{DBLP:journals/tvcg/ChenXR18,Kwon2020DPVisVA}, triangles~\cite[IV]{DBLP:journals/tvcg/WangPSSRMMS09}, rectangles~\cite[IV]{DBLP:journals/tvcg/UngerDSL18}, squares~\cite[IV]{DBLP:journals/tvcg/VrotsouN19,DBLP:journals/tvcg/LiuWDHWW17}, and other shapes~\cite[PM]{DBLP:journals/is/LeoniMA15}. Each event sequences is shown as a separate line. Some of these visualizations can highlight a particular event type in all sequences~\cite[IV]{DBLP:journals/tvcg/VrotsouN19} or align events according to one specific event type~\cite[IV]{DBLP:journals/tvcg/LiuWDHWW17,DBLP:journals/tvcg/WangPSSRMMS09}. These visualizations show from as few as five~\cite[IV]{DBLP:journals/tvcg/WangPSSRMMS09} to hundreds~\cite[IV]{DBLP:journals/tvcg/LiuWDHWW17} of event sequences in one view. Representations in this category convey also patterns of aggregated event sequence data. Nguyen et al.~\cite[IV]{DBLP:journals/tvcg/NguyenTAATZ19} use visual highlighting to show aggregations. Chen et al.~\cite[IV]{DBLP:journals/tvcg/ChenYPCSPQ20} distinguish different event types using different shapes.

A second group of {\chartA} visualizations display one or more event sequences using visual elements. Long sequences are split into several rows and glyphs are connected with lines~\cite[IV]{DBLP:journals/tvcg/GuoXZGZC18,DBLP:journals/tvcg/LawLMB19,DBLP:journals/tvcg/GuoJGDZC19,DBLP:journals/corr/abs-2009-00219}. This design is used to show details of a concrete event sequence for analysis. This representation can also be used for aggregated event sequences. MAQUI~\cite[IV]{DBLP:journals/tvcg/LawLMB19} shows the major patterns discovered from event sequence data.
Capper et al.~\cite[IV]{DBLP:journals/tvcg/CappersW18} present a visualization system for individual event sequences and aggregated event sequences, using a sequence of annotated rectangles capturing rules discovered from event sequences.
Chen et al.~\cite[IV]{DBLP:journals/tvcg/ChenXR18} introduced a visualization system that uses a {\chartA} representation. They use a sequence of colored rectangles to represent the aggregated event sequences. This visualization design supports interactive exploration of details related to specific patterns. These patterns are displayed as taller rectangles and aligned below the original aggregated sequence.

\paragraph{B. {\chartB}}

\begin{wrapfigure}{l}{0.25\textwidth}
	\centering
	\includegraphics[width=0.25\textwidth]{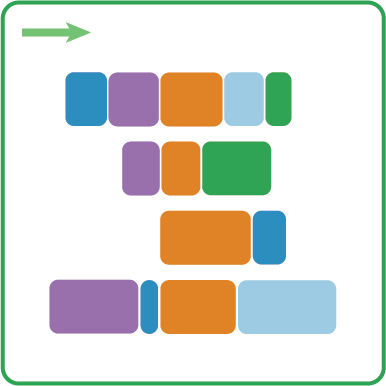}
	\caption{{\chartB} visualization}
	\label{fig-chart-B}
\end{wrapfigure}

{\chartB} visualizations map event sequences to the visual items in such a way that the size of the elements are proportional to characteristics such as the event duration (see~\Cref{fig-chart-B}). 

Loorak et al.~\cite[IV]{DBLP:journals/tvcg/LoorakPKHC16,DBLP:journals/tvcg/UngerDSL18} present a vertical design of {\chartB} visualization of event sequences and their duration. These sequences are arranged along the x-axis and the y-axis showing the duration of individual events in event sequence. Xu et al.~\cite[IV]{DBLP:journals/tvcg/XuMR017} describe a radial representation, arranging the event sequence around a circle. Rosenthal et al.~\cite[IV]{DBLP:journals/cgf/RosenthalPMO13} use a one-dimensional {\chartB} overview of event sequence data. The color of the rectangles indicate the event type and size shows the number of aggregated events.

\paragraph{C. {\chartC}}

Bar charts visualize data of different categories as \begin{wrapfigure}[7]{l}{0.25\textwidth}
	\centering
	\includegraphics[width=0.25\textwidth]{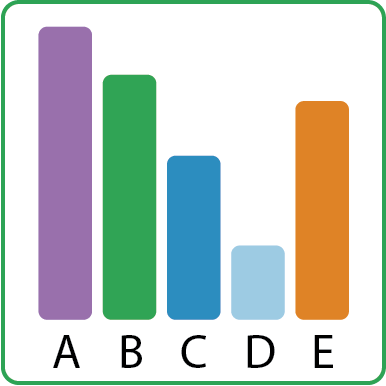}
	\caption{{\chartC} visualization}
	\label{fig-chart-C}
\end{wrapfigure} rectangular visual elements, where width is fixed and height represents some quantitative value (see~\Cref{fig-chart-C}). Stacked bar charts offer a comparison between different subcategories of data. 

Event sequence data visualization systems use bar chart views to present some distributional aspects of the data.
Some works use bar charts to show the distribution of gender~\cite[IV]{DBLP:journals/tvcg/KrausePS16,DBLP:journals/tvcg/GotzZWSB20,DBLP:journals/corr/abs-2009-00219}, race of participants~\cite[IV]{DBLP:journals/tvcg/GotzZWSB20}, or event frequencies~\cite[IV]{DBLP:journals/tvcg/GotzZWSB20,DBLP:journals/corr/abs-2009-00219,DBLP:journals/cgf/DortmontEW19,DBLP:journals/tvcg/XuMR017}. Kwon et al.~\cite[IV]{Kwon2020DPVisVA} provide bar charts to summarize various other characteristics of the event sequence data.
Agarwal et al.~\cite[IV]{DBLP:journals/cgf/AgarwalB20} give an overview of the number of events per a specified time ranges. Nguyen et al.~\cite[IV]{DBLP:journals/tvcg/NguyenTAATZ19} use the bar chart with a chart for each sequence. One bar represents two variables, one captured by its height and the other by its color.
Suriadi et al.~\cite[PM]{DBLP:journals/dss/SuriadiWXAH17} use bar charts to show the average length of queue before each event in the sequences. Low et al.~\cite[PM]{DBLP:journals/is/LowAHWW17} show the difference in resource utilization between two event sequence datasets. Bose et al.~\cite{DBLP:journals/is/BoseA12} uses bar charts as a additional view, in which each bar corresponding to one event and the extent its traces are aligned with a predefined model.

\subsubsection{Other Chart Types}
\paragraph{D. Histogram} A histogram is a visualization of data that is based on the bar chart. Instead of displaying bars for a categorical value, the histogram replaces them with ranges derived from a quantitative values.

Wang et al.~\cite[IV]{DBLP:journals/tvcg/WangPSSRMMS09} use histograms and stacked histograms to aggregate and show the distribution of event types on a daily basis. Similarly, the authors of~\cite[IV]{DBLP:journals/tvcg/ChenYPCSPQ20,DBLP:journals/tvcg/KrausePS16} use histograms to show the distribution of outcomes of an event sequence. Gotz et al.~\cite[IV]{DBLP:journals/tvcg/GotzZWSB20} and Jin et al.~\cite[IV]{DBLP:journals/corr/abs-2009-00219} use histograms to display age distribution in the data.
Nguyen et al.~\cite[IV]{DBLP:journals/tvcg/NguyenBJKBGMB21} utilize modified histograms to show the frequency of different process variants. These modified histograms use a second x-axis with another variable connecting the two axes with shadowed lines.

\paragraph{E. Line Chart} Line chart displays the series of data points connected by a line in a two-dimensional graph. It shows a change of a variable over time. 
Xie et al.~\cite[IV]{DBLP:journals/corr/abs-2009-02464} use line charts to display statistics of event sequences derived from football data. Jo et al.~\cite[IV]{DBLP:journals/tvcg/JoHPKS14} use a line chart to represent outcome changes per group of event sequences. They color slopes of this line chart use hue and saturation for an increase or decrease of value.

\paragraph{F. Scatter plot} Scatter plots show the distribution of values of a two-dimensional numerical variable. These values are plotted on the x- and y-axes.
Nguyen et al.~\cite[IV]{DBLP:journals/tvcg/NguyenBJKBGMB21} show a scatter plot of how two attributes correlate. In this way, they compare exclusive and inclusive run-time of event sequences from parallel program executions. 
Gotz et al.~\cite[IV]{DBLP:journals/tvcg/GotzZWSB20} present a so-called scatter-and-focus visualization. This interactive visualization uses a green-red color scale for differentiating the points and for choosing the focus points for further analysis. 
Chen et al.~\cite[IV]{DBLP:journals/tvcg/ChenYPCSPQ20} use a scatter plot as a projection view of their visualization. They display the event sequences aggregated and sorted by similarity with respect to a predefined time period.

\paragraph{G. Pie chart} Pie charts are charts that slice a circle for showing proportions taken by different groups.
Bolt et al.~\cite[PM]{DBLP:journals/is/BoltLA18} use pie charts to present the results of a classification algorithm for detection of differences between behavior and business rules. Liu et al.~\cite[PM]{DBLP:journals/tkde/LiuXPGX17} show a pie chart of the aggregated distribution of different causes of anomaly.

\paragraph{H. Heat map} Heat maps visualize quantities of data using a color scale projected onto a two-dimensional space. 
Isaacs et al.~\cite[IV]{DBLP:journals/tvcg/IsaacsBJGBSH14} employ the heat maps to visualize program execution traces. They arrange the event sequences in rows and their time in columns. The color scheme from light blue to red identifies the delay of execution of certain events. Lie et al.~\cite[PM]{DBLP:journals/tkde/LiuXPGX17} use a heat map to highlight the columns in a matrix according to the extent of abnormal behavior present in corresponding event sequences. 

\paragraph{I. Map} This visualization type positions visual elements on a two-dimensional map.
Leoni et al.~\cite[PM]{DBLP:journals/tvcg/LawLMB19} display individual events of event sequences on a city map. They use several glyph types (rectangles, circles) to distinguish event types.
 
\paragraph{J. Calendar} This visual representation utilizes a calendar of dates in the month to present information. 
Xu et al.~\cite[IV]{DBLP:journals/tvcg/XuMR017} use a calendar view to represent the number of faults for each day. They color  squares scaled by color lightness.

\paragraph{K. Density Plot} A density plot is a smoothed version of a histogram that is used to show how the numerical variable is distributed. It is used as a building block among others by Xu et al.~\cite[IV]{DBLP:journals/tvcg/XuMR017}, Wu et al.~\cite[IV]{DBLP:journals/tvcg/WuLYLW14}, and Dortmont et al.~\cite[IV]{DBLP:journals/cgf/DortmontEW19}.

\paragraph{L. Box Plot} A Box plot is a visualization method to aggregate and display a set of numerical values. This representation displays data through quartiles, drawing a rectangle showing second and third quartiles with lines extending further to show lower and upper quartiles. Bernard et al.~\cite[IV]{DBLP:journals/tvcg/BernardSKR19} use it as part of their visualizations.

\subsubsection{Compositions of Chart Types}
\noindent
A key property of technology is its compositional nature~\cite{arthur2009nature}. Also visualization techniques inherit this compositionality. Here, we describe works that integrate different chart-based techniques in one visualization view. Several authors use a combination of {\chartA} visualizations together with other chart-based visualizations. 

Chen et al.~\cite{DBLP:journals/tvcg/ChenYPCSPQ20} display individual event sequences as sequences of glyphs as in \emph{{\chartA}} overlaying with \emph{line charts} to show how quantities change with progress. Xie et al.~\cite[IV]{DBLP:journals/corr/abs-2009-02464} use a~\emph{map} visualization to position elements of a~\emph{\chartA}. They use the map of a football pitch to show the sequence of passes that lead to a certain situation (e.g. a goal).  

Low et al.~\cite[PM]{DBLP:journals/is/LowAHWW17} combine a~\emph{\chartA} visualization to show which activities have a time shift, and a \emph{diverging bar chart} to display the aggregated time shift between two event log datasets. Xie et al.~\cite{DBLP:journals/corr/abs-2009-02464} use a combination of~\emph{\chartA} and~\emph{\chartC} visualizations. They use the bar charts to show the aggregate current state of a system, while on each bar an event can be displayed as a circle to show individual event sequences. Isaacs et al.~\cite[IV]{DBLP:journals/tvcg/IsaacsBJGBSH14} use a combination of~\emph{\chartA},~\emph{\chartC} and~\emph{heatmaps} in their visualizations. The individual events are arranged by rows and columns as rectangles that are colored by hue and saturation to show how late events are due to some predetermined criteria. Additionally, at the bottom of this view, the bar chart shows the overview of the whole event sequence while the main view only shows the chosen part.

Richter et al.~\cite[PM]{DBLP:journals/is/RichterS19} and Zeng et al.~\cite[IV]{DBLP:journals/tvcg/ZengWWWLEQ20} use the~\emph{\chartB} to show individual event sequences next to a~\emph{line chart} to show how additional quantitative characteristic changes with time of the sequence.
Richter et al.~\cite[PM]{DBLP:journals/is/RichterS19} also combines \emph{line chart} and \emph{scatter plot} using the same x-axis to display two different characteristics for each transitions between two event types. The scatter plot shows the duration of these transitions and a line chart for identifying when the changes in behavior occur. 

Xu et al.~\cite[IV]{DBLP:journals/tvcg/XuMR017} use a~\emph{\sankeyA} visualization arranged vertically. For each event type, a small density plot is displayed showing the occurrence distribution of this event in the data.
Chen et al.~\cite[IV]{DBLP:journals/tvcg/ChenAAABKNTT20} use a \emph{scatter plot} to arrange \emph{pie charts} on the two dimensional space.
Bernard et al.~\cite[IV]{DBLP:journals/tvcg/BernardSKR19} group several different charts into a group, and a sequence of instances of this group of charts to signify how the characteristic of a group of event sequences changes with time. In this design the~\emph{bar charts},~\emph{box plots},~\emph{pie charts},~\emph{matrix views} are combined into a group to show the characteristics. This system was designed for a medical system for monitoring to analyse cohorts of patients.

\subsection{Discussion}
\noindent
We analyzed contributions from the fields of process mining and information visualization using the ESeVis framework. We observe that both fields differ in terms of their emphasis on sequence representations and model representations. 

Contributions focusing on \emph{instance representations} are dominated by the field of information visualization. A larger share of these works are in the category of timeline-based visualizations and in the category of other chart types. The articles of our review include few works that build on matrix-based, hierarchy-based, or sankey-based visualizations of event sequences. There are very few contributions from process mining research on instance representation. 
Contributions on \emph{model representations} are more balanced between both research fields. A larger share of these works are sankey-based or in the category of other chart types. There are some hierarchy-based, some timeline-based, and few matrix-based works. 
Process mining contributions are to a larger share related to sankey diagrams and the category of other chart-based visualizations. 

These findings highlight some strengths of the \emph{process mining} field.
They reflect the fact that process mining research puts a strong emphasis on discovery algorithms that generate process models such as variants of node-link diagrams or Petri nets from event sequence data~\cite{DBLP:books/sp/Aalst16}.
We observe that this focus on designing new algorithms integrates into a fertile discourse on which knowledge can be extracted from event sequence data.
The findings also point to strengths of the \emph{information visualization} field. 
This field puts a strong emphasis on the creative development of powerful visual interfaces to interact with complex analysis algorithms, often designed for the specific requirements of a given application domain. Formal notations like Petri nets are hardly considered.
Our findings underline the potential for stronger synergies of research on process mining and information visualization.

Our review extends insights from previous review articles.
Several frameworks on corresponding review articles have been written in the field of \emph{information visualization}.
McNabb et al.~\cite{DBLP:journals/cgf/McNabbL17} introduce a survey of surveys of information visualization. They categorize the field into topic clusters using a 2-dimensional classification schema. For the first dimension, the phase of information visualization pipeline~\cite{DBLP:books/daglib/0098275} is used. In the second dimension, the survey papers are categorized by the subject of the study~\cite{DBLP:books/daglib/0098275} (e.g. graphs and networks vs. geospace). While some of the categories mention the use of data including events (geospace+time, data-centric surveys), this survey of surveys does not point to an existing survey that reviews event sequence data visualizations.

Aigner et al.~\citep{DBLP:series/hci/AignerMST11,DBLP:journals/cg/AignerMMST07} focus their survey on time-oriented data visualizations. The survey in~\cite{DBLP:journals/cg/AignerMMST07} categorizes visualization designs by three categories: time, data, and representation. In their taxonomy, the notion of event data is introduced as part of the data category, showing that events represent one level of data abstraction. 
The recent survey, by Gue et al.~\cite{DBLP:journals/corr/abs-2006-14291} introduces a summary of visual analysis approaches for event sequence data. They propose a taxonomy based on research articles in the field of information visualization. They categorize visualizations into design space clusters: data scales, analysis techniques, visual representation, and interactions. Another focus of the survey is identifying and clustering the papers according to visual analytics tasks. Our survey adopted five visual representation categories from~\cite{DBLP:journals/corr/abs-2006-14291}. Within each of these categories, we extended the categorization into subcategories bringing a fine-granular distinction in visualization designs. 
None of these review articles explicitly considers contributions from process mining. 
%

There are also several review articles on \emph{process mining}. Their focus is on the comparison of process discovery algorithms, while the visual representations of discovery results are usually limited to assumed options. 
Weerdt et al.~\cite{DBLP:journals/is/WeerdtBVB12} present a multi-dimensional quality assessment of process discovery algorithms. Their quality assessment focuses on distinguishing process discovery algorithm types, and typical challenges of discovery (e.g. a problem of incorrectly detecting a loop in a model). They evaluate the results of process discovery for each of the surveyed approaches after conversion of the result to Petri nets representation. Visualization is not considered.
A more recent survey article, by Augusto et al.~\cite{DBLP:journals/tkde/AugustoCDRMMMS19} explicitly clusters articles by model languages produced by their process discovery algorithms in categories such as process trees, casual nets, BPMN models, state machines. To assess the quality of results, this survey still converts these representations into Petri nets, and does not focus on visualizations of these models.
Only recently, the effective visualization of process mining results has become a subject of research~\cite{DBLP:conf/bpm/MendlingDM21}, building on insights on the cognitive effectiveness of diagrams~\cite{malinova2021cognitive}.
Contributions from information visualization provide concepts to inform this emerging research.

%% file: tables/4-matrix.tex
\rowcolors{3}{white}{gray!12.5}
\begin{tabular}{p{0.1\linewidth} c c }
	\toprule
	\cellcolor{white} & Matrix & Extended matrix  \\
	\cellcolor{white} & 
	\includegraphics{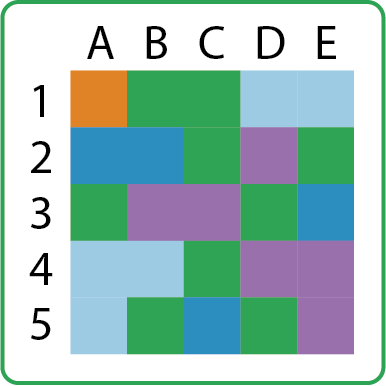} & 
	\includegraphics{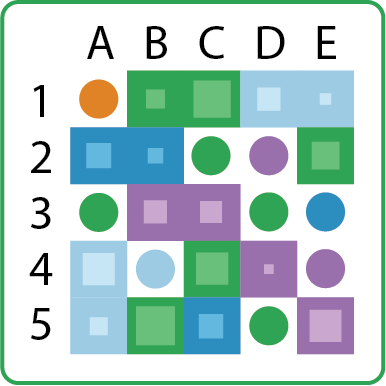}   \\  \midrule
	
	\acrshort{tvcg} &  \thead{Chen et al.~\cite{DBLP:journals/tvcg/ChenAAABKNTT20} \\ Kwon et al.~\cite[IV]{Kwon2020DPVisVA}}& \thead{Loorak et al.~\cite{DBLP:journals/tvcg/LoorakPKHC16} \\ Zhuochen et al.~\cite{DBLP:journals/corr/abs-2009-00219} \\ Xiao et al.~\cite{DBLP:journals/corr/abs-2009-02464}} \\
	
	\acrshort{iv} & & \\
	
	\acrshort{cgf} & &  \\
	
	\acrshort{is} &  Bose et al.~\cite{DBLP:journals/is/BoseA12} & \\
	
	\acrshort{tkde} & & \\
	
	\acrshort{dss} && \\
	
	\bottomrule

\end{tabular} 

%% file: tables/4-timeline.tex
\rowcolors{3}{white}{gray!12.5}
\begin{tabular}{p{0.1\linewidth} r r r r r }
	\toprule
	\cellcolor{white} & Fixed & Duration & \thead{Converging-\\Diverging} & Evolution & Combination \\
	\cellcolor{white} & 
	\includegraphics{graphics/sec4/timeline/1-fixed} & 
	\includegraphics{graphics/sec4/timeline/2-duration.png}  & \includegraphics{graphics/sec4/timeline/3-converging-diverging.png}  &  \includegraphics{graphics/sec4/timeline/4-evolution.png}  &
	\includegraphics{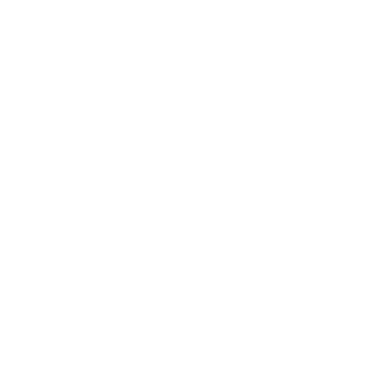}
	\\  \midrule
	\acrshort{tvcg} & \thead{Chen et al.~\cite{DBLP:journals/tvcg/ChenXR18} \\ Kwon et al.~\cite{Kwon2020DPVisVA} \\ Monroe et al.~\cite{DBLP:journals/tvcg/MonroeLLPS13} \\ Fulda et al.~\cite{DBLP:journals/tvcg/FuldaBM16}} & \thead{Nguyen et al.~\cite{DBLP:journals/tvcg/NguyenTAATZ19}\\Vrotsou et al.~\cite{DBLP:journals/tvcg/VrotsouN19}\\Zeng et al.~\cite{DBLP:journals/tvcg/ZengWWWLEQ20} \\Vrotsou et al.~\cite{DBLP:journals/tvcg/VrotsouJC09}} &  \thead{Xu et al.~\cite{DBLP:journals/tvcg/XuMR017} \\ Liu et al.~\cite{DBLP:journals/tvcg/LiuWWLL13} \\ Kwon et al.~\cite{Kwon2020DPVisVA} \\ Baumgartl et al.~\cite{DBLP:journals/corr/abs-2008-09552}} & \thead{Wu et al.~\cite{DBLP:journals/tvcg/WuLYLW14}}  & \thead{Fulda et al.~\cite{DBLP:journals/tvcg/FuldaBM16} \\ Monroe et al.~\cite{DBLP:journals/tvcg/MonroeLLPS13} \\Jo et al.~\cite{DBLP:journals/tvcg/JoHPKS14}} \\
	
	\acrshort{iv} & & \thead{Vrotsou et al. ~\cite{DBLP:journals/ivs/VrotsouYC14}}& & & \\
	
	\acrshort{cgf} &  \thead{Leite et al.~\cite{DBLP:journals/cgf/LeiteGMGK20} \\  Dortmont et al.~\cite{DBLP:journals/cgf/DortmontEW19} } & \thead{Han et al.~\cite{DBLP:journals/cgf/HanRDAS15} \\ Rosenthal et al.~\cite{DBLP:journals/cgf/RosenthalPMO13}}& \thead{Reda et al.~\cite{DBLP:journals/cgf/RedaTJLB11}}  & \thead{Sung et al.~\cite{DBLP:journals/cgf/SungHSCLW17} \\ Dortmont et al.~\cite{DBLP:journals/cgf/DortmontEW19}} & \thead{Sung et al.~\cite{DBLP:journals/cgf/SungHSCLW17} \\ Liu et al.~\cite{DBLP:journals/cgf/LiuKDGHW17}}\\
	
	\acrshort{is} & Bose et al.~\cite{DBLP:journals/is/BoseA12} & \thead{Low et al.~\cite{DBLP:journals/is/LowAHWW17} \\Richter et al.~\cite{DBLP:journals/is/RichterS19}}& & & \\
	
	\acrshort{tkde} & & & & & \\
	
	\acrshort{dss} & \thead{Leoni et al.~\cite{DBLP:journals/dss/LeoniAAH12}} & & & \thead{Suriadi et al.~\cite{DBLP:journals/dss/SuriadiOAH15}\\Suriadi et al.~\cite{DBLP:journals/dss/SuriadiWXAH17}} & \\
	
	\bottomrule
\end{tabular} 

%% file: tables/4-hierarchy.tex
\rowcolors{3}{white}{gray!12.5}
\begin{tabular}{p{0.1\linewidth} c c c }
	\toprule
	\cellcolor{white} & Node-link & Flow & Treemap \\
	\cellcolor{white} & 
	\includegraphics{graphics/sec4/hierarchy/1-nodelink} & 
	\includegraphics{graphics/sec4/hierarchy/2-flow}  & \includegraphics{graphics/sec4/hierarchy/3-treemap} \\  \midrule
	\acrshort{tvcg} & \thead{Law et al.~\cite{DBLP:journals/tvcg/LawLMB19} \\ Vrotsou et al.~\cite{DBLP:journals/tvcg/VrotsouJC09} \\ Vrotsou et al.~\cite{DBLP:journals/tvcg/VrotsouN19} \\ Krause et al.~\cite{DBLP:journals/tvcg/KrausePS16}}  & \thead{Wongsuphasawat et al.~\cite{DBLP:journals/tvcg/WongsuphasawatG12} \\ Guo et al.~\cite{DBLP:journals/tvcg/GuoXZGZC18} \\ Jang et al.~\cite{DBLP:journals/tvcg/JangER16} \\ Liu et al.~\cite{DBLP:journals/tvcg/LiuWDHWW17} \\ Law et al.~\cite{DBLP:journals/tvcg/LawLMB19} \\ Guo et al.~\cite{DBLP:journals/tvcg/GuoJGDZC19}} & Jang et al.~\cite{DBLP:journals/tvcg/JangER16} \\
	
	\acrshort{iv} & & & \\
	
	\acrshort{cgf} & \thead{Leite et al.~\cite{DBLP:journals/cgf/LeiteGMGK20}  \\ Liu et al.~\cite{DBLP:journals/cgf/LiuKDGHW17}} & Liu et al.~\cite{DBLP:journals/cgf/LiuKDGHW17} & \\
	
	\acrshort{is} & Bose et al.~\cite{DBLP:journals/is/BoseA12} & & \\
	
	\acrshort{tkde} & & & \\
	
	\acrshort{dss} & \thead{Song et al.~\cite{DBLP:journals/dss/SongA08} \\ Leoni et al.~\cite{DBLP:journals/dss/LeoniAAH12}} & & \\
	
	\bottomrule

\end{tabular}

%% file: tables/4-sankey.tex
\rowcolors{3}{white}{gray!12.5}
\begin{tabular}{p{0.1\linewidth} r r r r r }
	\toprule
	\cellcolor{white} & Node-link & \thead{Extended \\ node-link} & Flow & \thead{Converging-\\Diverging} & Chord \\
	\cellcolor{white} & 
	\includegraphics{graphics/sec4/sankey/1-node-link.png} & 
	\includegraphics{graphics/sec4/sankey/2-ex-node-link.png}  & \includegraphics{graphics/sec4/sankey/3-flow.png}  &  \includegraphics{graphics/sec4/sankey/4-converging.png}  &
	\includegraphics{graphics/sec4/sankey/5-chord.png}
	\\  \midrule
	\acrshort{tvcg} & \thead{Baumgartl et al.~\cite{DBLP:journals/corr/abs-2008-09552} \\ Chen et al.~\cite{DBLP:journals/tvcg/ChenYPCSPQ20}} & \thead{Jang et al.~\cite{DBLP:journals/tvcg/JangER16} \\ Zeng et al.~\cite{DBLP:journals/tvcg/ZengWWWLEQ20}} & \thead{Nguyen et al.~\cite{DBLP:journals/tvcg/NguyenBJKBGMB21} \\ Gotz et al.~\cite{DBLP:journals/tvcg/GotzZWSB20} \\ Jin et al.~\cite{DBLP:journals/corr/abs-2009-00219} \\ Cui et al.~\cite{DBLP:journals/tvcg/CuiLWW14} \\ Shi et al.~\cite{DBLP:journals/tvcg/ShiWLZQ14}\\ Wu et al.~\cite{DBLP:journals/tvcg/WuLYLW14} \\ Zeng et al.~\cite{DBLP:journals/tvcg/ZengWWWLEQ20} \\ Qi et al.~\cite{DBLP:journals/tvcg/QiBWWW20} \\ Guo et al.~\cite{DBLP:journals/tvcg/GuoJGDZC19} } & Bartolomeo et al.~\cite{DBLP:journals/tvcg/BartolomeoZSD21} & Chen et al.~\cite{DBLP:journals/tvcg/ChenYPCSPQ20} \\
	
	\acrshort{iv} & && & Vrotsou et al.~\cite{DBLP:journals/ivs/VrotsouYC14} & \\
	
	\acrshort{cgf} & & Leite et al.~\cite[IV]{DBLP:journals/cgf/LeiteGMGK20}  & & \thead{Chou et al.~\cite{DBLP:journals/cgf/ChouWM19} \\ Agarwal et al.~\cite{DBLP:journals/cgf/AgarwalB20}} & \\
	
	\acrshort{is} & \thead{Bose et al.~\cite{DBLP:journals/is/BoseA12} \\ Winter et al.~\cite{DBLP:journals/is/WinterSR20}} & \thead{Munoz{-}Gama et al.~\cite{DBLP:journals/is/Munoz-GamaCA14} \\ Leoni et al.~\cite{DBLP:journals/is/LeoniMA15} \\ Bolt et al.~\cite{DBLP:journals/is/BoltLA18} \\ Low et al.~\cite{DBLP:journals/is/LowAHWW17} \\  Knuplesch et al.~\cite{DBLP:journals/is/KnupleschRK17} \\ Dongen et al.~\cite{DBLP:journals/corr/abs-2011-11551}}  & & & \\
	
	\acrshort{tkde} & & & & & \\
	
	\acrshort{dss} & Song et al.~\cite{DBLP:journals/dss/SongA08} & \thead{Song et al.~\cite{DBLP:journals/dss/SongA08} \\ Wynn et al.~\cite{DBLP:journals/dss/WynnPXHBPA17}} & & & \\
	
	\bottomrule

\end{tabular}

%% file: tables/4-chart.tex
\rowcolors{3}{white}{gray!12.5}
\begin{tabular}{p{0.1\linewidth} r r r r r }
	\toprule
	\cellcolor{white} & Sequence & \thead{Sequence-\\Duration} & Bar-chart & Other chart & Composition \\
	\cellcolor{white} & 
	\includegraphics{graphics/sec4/chart/1-sequence.png} & 
	\includegraphics{graphics/sec4/chart/2-seq-duration.png}  & \includegraphics{graphics/sec4/chart/3-bar.png} &  \includegraphics{graphics/sec4/timeline/combinations.png}  &
	\includegraphics{graphics/sec4/timeline/combinations.png}
	\\  \midrule
	\acrshort{tvcg} &\thead{Kwon et al.~\cite{Kwon2020DPVisVA} \\ Wang et al.~\cite{DBLP:journals/tvcg/WangPSSRMMS09} \\ Unger et al.~\cite{DBLP:journals/tvcg/UngerDSL18} \\ Liu et al.~\cite{DBLP:journals/tvcg/LiuWDHWW17} \\ Vrotsou et al.~\cite{DBLP:journals/tvcg/VrotsouN19}\\ Nguyen et al.~\cite{DBLP:journals/tvcg/NguyenTAATZ19} \\ Chen et al.~\cite{DBLP:journals/tvcg/ChenYPCSPQ20} \\ Guo et al.~\cite{DBLP:journals/tvcg/GuoXZGZC18} \\ Guo et al.~\cite{DBLP:journals/tvcg/GuoJGDZC19} \\ Jin et al.~\cite{DBLP:journals/corr/abs-2009-00219} \\ Law et al.~\cite{DBLP:journals/tvcg/LawLMB19} \\ Cappers et al.~\cite{DBLP:journals/tvcg/CappersW18} \\ Chen et al.~\cite{DBLP:journals/tvcg/ChenXR18}} & \thead{Loorak et al.~\cite{DBLP:journals/tvcg/LoorakPKHC16} \\ Unger et al.~\cite{DBLP:journals/tvcg/UngerDSL18} \\ Xu et al.~\cite{DBLP:journals/tvcg/XuMR017}} &\thead{Krause et al.~\cite{DBLP:journals/tvcg/KrausePS16} \\Xu et al.~\cite{DBLP:journals/tvcg/XuMR017} \\Jin et al.~\cite{DBLP:journals/corr/abs-2009-00219} \\Kwon et al.~\cite[IV]{Kwon2020DPVisVA} \\Nguyen et al.~\cite[IV]{DBLP:journals/tvcg/NguyenTAATZ19}}&\thead{Wang et al.~\cite{DBLP:journals/tvcg/WangPSSRMMS09} \\ Krause et al.~\cite{DBLP:journals/tvcg/KrausePS16} \\ Gotz et al.~\cite{DBLP:journals/tvcg/GotzZWSB20} \\ Jin et al.~\cite{DBLP:journals/corr/abs-2009-00219} \\ Nguyen et al.~\cite{DBLP:journals/tvcg/NguyenBJKBGMB21} \\ Xie et al.~\cite{DBLP:journals/corr/abs-2009-02464} \\ Jo et al.~\cite{DBLP:journals/tvcg/JoHPKS14} \\ Chen et al.~\cite{DBLP:journals/tvcg/ChenYPCSPQ20} \\ Isaacs et al.~\cite[IV]{DBLP:journals/tvcg/IsaacsBJGBSH14} \\ Leoni et al.~\cite[PM]{DBLP:journals/tvcg/LawLMB19} \\ Xu et al.~\cite[IV]{DBLP:journals/tvcg/XuMR017}}&\thead{Chen et al.~\cite{DBLP:journals/tvcg/ChenYPCSPQ20} \\ Xie et al.~\cite{DBLP:journals/corr/abs-2009-02464} \\ Isaacs et al.~\cite{DBLP:journals/tvcg/IsaacsBJGBSH14} \\ Zeng et al.~\cite{DBLP:journals/tvcg/ZengWWWLEQ20} \\ Xu et al.~\cite{DBLP:journals/tvcg/XuMR017} \\ Chen et al.~\cite{DBLP:journals/tvcg/ChenAAABKNTT20} \\ Bernard~\cite{DBLP:journals/tvcg/BernardSKR19}}\\
	
	\acrshort{iv} & &&&& \\
	
	\acrshort{cgf} &&Rosenthal et al.~\cite{DBLP:journals/cgf/RosenthalPMO13}&\thead{Dortmont et al.~\cite{DBLP:journals/cgf/DortmontEW19} \\
		Agarwal et al.~\cite{DBLP:journals/cgf/AgarwalB20}}&&\\
	
	\acrshort{is} &Leoni et al.~\cite[PM]{DBLP:journals/is/LeoniMA15} && \thead{Low et al.~\cite{DBLP:journals/is/LowAHWW17} \\ Bose et al.~\cite{DBLP:journals/is/BoseA12}}& Bolt et al.~\cite{DBLP:journals/is/BoltLA18} & \thead{Richter et al.~\cite{DBLP:journals/is/RichterS19}  \\ Low et al.~\cite{DBLP:journals/is/LowAHWW17}} \\
	
	\acrshort{tkde} & & & & Lie et al.~\cite{DBLP:journals/tkde/LiuXPGX17} & \\
	
	\acrshort{dss} & &&Suriadi et al.~\cite[PM]{DBLP:journals/dss/SuriadiWXAH17} & & \\
	
	\bottomrule
\end{tabular}

%% file: sections/conclusions.tex
\noindent 
In this paper, we have presented a review of visualization approaches for event sequence data. To this end, we have developed the ESeVis Framework for categorizing visualizations from information visualization and process mining fields. Our findings highlight strengths of the \emph{process mining} field in terms of generating formal models and the \emph{information visualization} of creatively developing of powerful visual interfaces to interact with complex analysis algorithms.
Our findings underline the potential for stronger synergies of research on process mining and information visualization.
Future research can utilize our work for identifying useful contributions from both fields, which will eventually help towards the integration of these fields.